\documentclass[prx,aps,showkeys,showpacs,footnoteinbib,onecolumn,
unsortedaddress]{revtex4-2}
\usepackage{graphicx} 
\usepackage{dcolumn} 
\usepackage{amssymb}

\usepackage{physics}
\usepackage{natbib}
\usepackage{xcolor}
\usepackage{soul}
\usepackage{siunitx}
\usepackage{float}
\usepackage{esdiff}
\usepackage{bbold}
\usepackage{mathtools}
\usepackage{todonotes}
\usepackage{hyperref}
\usepackage{amsmath,amssymb,epic,eepic}
\usepackage[binary-units]{siunitx}
\usepackage[utf8]{inputenc}
\usepackage{graphicx}
\usepackage{hyperref}
\usepackage[capitalise]{cleveref}
\usepackage{todonotes}
\usepackage{bbm}
\usepackage{braket}
\usepackage{fontawesome}

\usetikzlibrary{%
	shapes.geometric,calc,intersections,%
	decorations.markings,decorations.text,decorations,%
	patterns,decorations.pathmorphing,fpu,%
	decorations.pathreplacing,%
	positioning,automata,shapes.multipart,circuits,%
	circuits.ee, circuits.ee.IEC, shapes.gates.ee,%
	arrows.meta,arrows
}

\newcommand{\mytodo}[1]%
{{\todo[inline,backgroundcolor=blue!10!white]{#1}
}}
\newcommand{\me}{\mathrm{e}}
\newcommand{\mi}{\mathrm{i}}
\newcommand{\md}{\mathrm{d}}

\begin{document}

\title{Time-resolved sensing of electromagnetic fields with single-electron interferometry}

\author{H. Bartolomei$^{1}$}
\author{E. Frigerio$^{1}$}
\author{M. Ruelle$^{1}$}
\author{G. Rebora$^{2}$}
\author{Y. Jin$^{3}$}
\author{U. Gennser$^{3}$}
\author{A. Cavanna$^{3}$}
\author{E. Baudin$^{1}$}
\author{J.-M. Berroir$^{1}$}
\author{I. Safi$^{4}$}
\author{P. Degiovanni$^{2}$}
\author{G. C. \surname{M\'enard}$^{1}$}
\email{email: gerbold.menard@phys.ens.fr}
\author{G. F\`eve$^{1}$}
\email{email: gwendal.feve@phys.ens.fr}
\affiliation{$^{1}$ Laboratoire de Physique de l’Ecole normale sup\'erieure, ENS, Universit\'e PSL, CNRS, Sorbonne Universit\'e, Universit\'e Paris Cit\'e, F-75005 Paris, France}
\affiliation{$^{2}$ Univ Lyon, ENS de Lyon, Université Claude Bernard Lyon 1, CNRS, Laboratoire de Physique, F-69342 Lyon, France}
\affiliation{$^{3}$ Centre de Nanosciences et de Nanotechnologies (C2N), CNRS, Universit\'e Paris-Saclay, 91120 Palaiseau, France}
\affiliation{$^{4}$ Laboratoire de Physique des Solides-CNRS-UMR5802. University Paris-Saclay, B\^at.510, 91405 Orsay, France}

\begin{abstract}
Characterizing quantum states of the electromagnetic field at microwave frequencies requires fast
and sensitive detectors that can simultaneously probe the field time-dependent amplitude and its
quantum fluctuations. In this work, we demonstrate a quantum sensor that exploits the phase of
a single electron wavefunction, measured in an electronic Fabry-Perot interferometer, to detect a
classical time-dependent electric field. The time resolution, limited by the temporal width of the
electronic wavepacket, is a few tens of picoseconds. The interferometry technique provides a voltage
resolution of a few tens of microvolts, corresponding to a few microwave photons. Importantly, our
detector simultaneously probes the amplitude of the field from the phase of the measured interference
pattern and its fluctuations from the interference contrast. This capability paves the way for on-chip
detection of quantum radiation, such as squeezed or Fock states.
\end{abstract}


\date{\today}

\maketitle


\section*{Introduction}

In recent years, tremendous progress has been made in the field of electron quantum optics \cite{Bocquillon14,Bauerle18}, aiming at the generation and manipulation of electronic quantum states propagating in nano-conductors. Single electron sources \cite{levitov1996}, have been implemented \cite{Feve07, Leicht11, Hermelin11, Dubois13, Fletcher13}  and their coherence properties have been characterized from two-particle interferometry \cite{Ol'kovskaya08, Bocquillon13,Dubois13}. Tomography protocols for the reconstruction of single electron states have also been proposed \cite{Grenier11,Roussel21} and experimentally realized \cite{Jullien14,Bisognin19,fletcher2019}. In the meantime, various electronic interferometers have been demonstrated and studied \cite{Ji03, Roulleau08,Mcclure09}. In the context of electron quantum optics, interferometers can be used to characterize\cite{haack2011} and manipulate\cite{Yamamoto12} quantum electronic states , for the encoding and processing of quantum information \cite{Roussel21}, and for the readout of quantum entanglement \cite{Splettstoesser09, dasenbrook2015, Thibierge16}. Long confined to very pure GaAs/AlGaAs heterostructures, these techniques are now developing rapidly in other materials such as graphene \cite{Deprez21, Ronen21,Jo22,Assouline23}. This shows that the field has reached the level of maturity needed for the development of its applications in two main different directions: for the processing of quantum information encoded in electronic flying qubits \cite{Edlbauer22} and quantum sensing based on single electron wavefunctions \cite{Souquet24}. 

In  electron quantum optics, quantum sensing would exploit the quantum coherence of single-electron states for the detection of quantum objects, such as quantum states of the electromagnetic field. Quantum radiation can be characterized by the relatively high frequency of the electromagnetic field (GHz and beyond) and by a small number of photon excitations. The zero-point fluctuations of the voltage of a single mode of the electromagnetic field at a frequency $f$ can be written as $V_\mathrm{ZPF} \approx \frac{hf}{e} \sqrt{z}$ \cite{Supp}, where $z$ is the ratio of the characteristic impedance of the resonator to the resistance quantum. For a $50$ Ohms characteristic impedance, and $f=10$ GHz, $V_\mathrm{ZPF} \approx  \SI{2}{\micro\volt}$ (which can be increased by impedance matching techniques). The detection of a quantum field thus requires the use of fast (subnanosecond temporal resolution) and sensitive (microvolt voltage resolution) detectors. 

\begin{figure*}
    \begin{centering}
    \includegraphics[width = \textwidth]{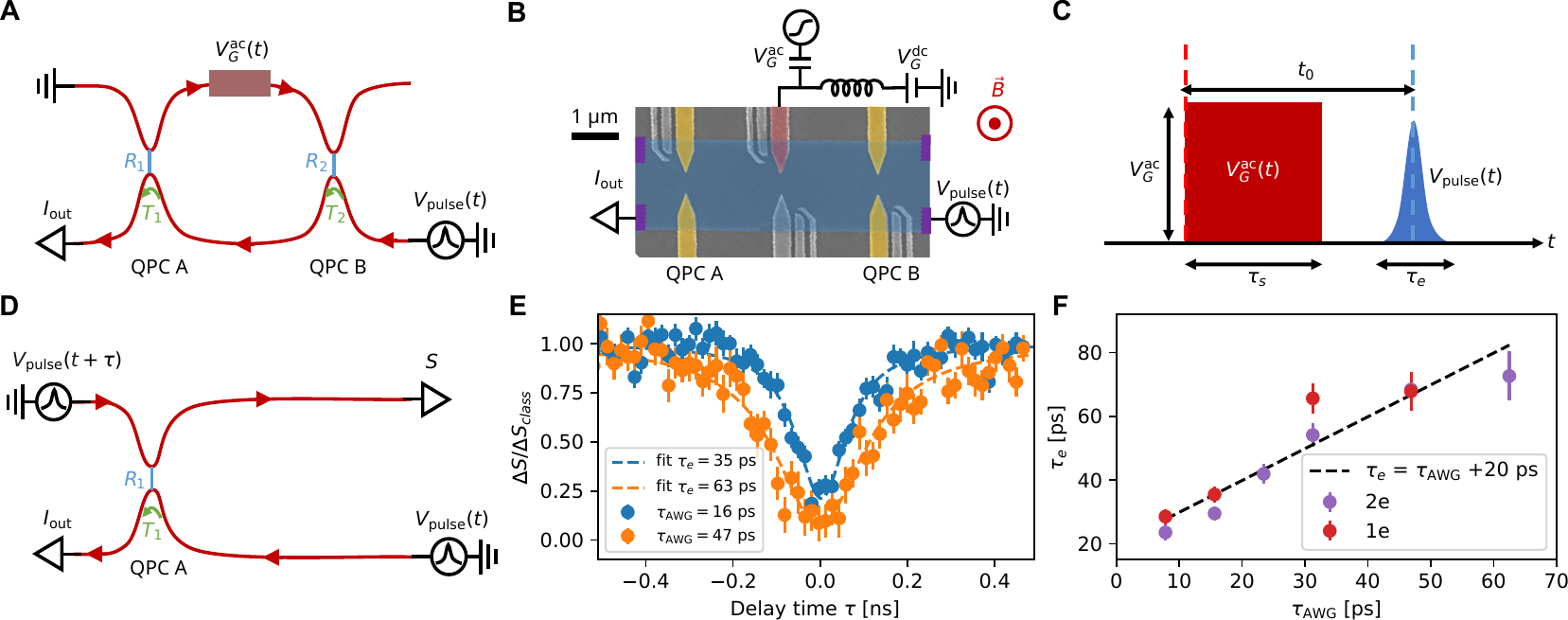}
    \caption{\textbf{Principle of the experiment.} (\textbf{A}) QPC A and QPC B are used for the partitioning of the outer edge channel with transmissions $T_i=1-R_i$ ($i=1,2$), defining an electronic Fabry-Perot interferometer (FPI). A single-electron pulse $V_\mathrm{pulse}(t)$ is sent through the bottom right branch of the interferometer. The square voltage $V_G^\mathrm{ac}(t)$ we probe is imposed on the central gate located on the upper arm of the FPI. The dc current $I_\mathrm{out}$ is measured at the output of the FPI at the bottom left of the schematics. (\textbf{B}) False-color electron microscopy image of the sample. QPCs 1 and 2 are indicated in yellow. The excitation gate (in red) is connected through a bias-tee to a dc and ac source such that we can send both a rf square excitation $V_G^\mathrm{ac}(t)$ and dc voltage $V_G^\mathrm{dc}$ . A magnetic field $\vec{B}$ perpendicular to the surface is applied to the sample. The 2DEG mesa is indicated in blue and Ohmic contacts are indicated in purple. All the other gates (in gray) are not used in this experiment and are left floating. (\textbf{C}) $V_\mathrm{pulse}(t)$ is a single electron Lorentzian pulse of width $\tau_e$. The square voltage $V_G^\mathrm{ac}(t)$ has a width $\tau_s$. The square voltage has a peak to peak amplitude $V_{G}^\mathrm{ac}$. (\textbf{D}) In a Hong-Ou-Mandel (HOM) configuration, we probe the width of the Lorentzian pulses by measuring the current noise coming out of QPC A. (\textbf{E}) Result of HOM experiment showing the amplitude of the noise (dots) as a function of the time difference between the two incoming Lorentzian pulses on QPC A for $\tau_\mathrm{AWG}=16$ (blue) and \SI{47}{\pico\second} (orange). A fit of the data (in dotted lines) shows that the actual time-widths at the level of the sample are 35 and \SI{63}{\pico\second}. (\textbf{F}) Measured width of the pulses as a function of the set time on the AWG for pulses containing one (red) or two (purple) electrons. We observe a linear dependence $\tau_e=\tau_\mathrm{AWG}+$~\SI{20}{\pico\second}.}
    \label{fig1}
    \end{centering}
\end{figure*}

In addition, quantum states of the electromagnetic field, such as Fock or squeezed vaccuum states, have a vanishing average field amplitude, such that all the information on the state is encoded in the field fluctuations.  This imposes a huge challenge for the development of quantum detectors that would be both fast and sensitive and would simultaneously probe the amplitude and the fluctuations of the electromagnetic field.  The short temporal width (a few picoseconds) of single electron states, which naturally points to the required short temporal resolution, has already been exploited for the picosecond sampling of a time-dependent voltage \cite{Johnson17}. This first demonstration of a sensing application using single electrons relied on the time modulation of the transmission probability through an energy selective potential barrier\cite{ubbelohde2015}. This method did not exploit the quantum coherence of single electron states, bringing limitations in terms of the sensitivity (of the order of a few hundred microvolts), but more importantly on the possibility to detect quantum states, as such method would only probe the amplitude of the detected voltage and be insensitive to quantum fluctuations. 

In this work, we exploit for the first time the quantum coherence of single electron states measured in an electronic Fabry-Perot interferometer (FPI) for the readout of a fast time-dependent voltage applied on a gate placed in one arm of the interferometer (see Fig. \ref{fig1}A). As already demonstrated in previous experiments \cite{Johnson17}, the short temporal width of single electron states brings a temporal resolution of a few tens of picoseconds. In addition, the  interferometry technique we use here has a sensitivity of a few tens of microvolts, which could be improved by optimizing the geometry of our  detector. More importantly, the quantum nature of our detection scheme, where the detected field is directly imprinted in the phase of the electronic wavefunction, is naturally suited for the detection of quantum radiation. The amplitude of the field is directly extracted from the amplitude of the phase shift of the measured interference pattern. The field fluctuations can also be directly extracted as a reduction of the interference contrast associated to the fluctuations of the electronic phase. Our method thus opens the way to the detection of non-classical states of the electromagnetic field, such as Fock or squeezed states \cite{Souquet24}. Finally, our method offers completely new possibilities of detection based on the engineering of the phase of single electron states with, for example, applications to the down-conversion of THz radiation. 

\section*{Sample and characterization of single electron pulses}

Our sample, shown on Fig. \ref{fig1}B, is a two-dimensional electron gas (GaAs/AlGaAs) of density $n_s = 1.2\times 10^{15}$~\SI{}{\per\meter\squared} and mobility $\mu = 1.8 \times 10^6 $~\SI{}{\centi\meter\squared\per\volt\per\second}, set at filling factor $\nu= 3$ by applying a perpendicular magnetic field $B =$~\SI{1.37}{T} \cite{Supp}. The interferometer is a FPI of height $H=2 \pm 0.2$~\SI{}{\micro\meter} (taking into account a depletion length of $0.25 \pm 0.1$~\SI{}{\micro\meter} on each side of the sample) and width $W=3.6 \pm 0.2$~\SI{}{\micro\meter}. The perimeter of the FPI is $L=2 \times (H+W)=11.2 \pm 0.8$~\SI{}{\micro\meter} and its area is $A=7.2 \pm 0.8~$\SI{}{\micro\meter\squared}. Two quantum point contacts (QPCs) (in yellow on Fig. \ref{fig1}B) are used to partition the outer channel at $\nu=3$ with a transmission probability $T_i$ ($i=1,2$) and reflection probability $R_i=1-T_i$ (see Fig. \ref{fig1}A). This implies that single electron interferometry occurs on the outer channel of $\nu=3$ while the two inner channels (not represented on Fig. \ref{fig1}A) are fully reflected within the cavity. 

The red gate on Fig. \ref{fig1}B is used as a plunger to tune the interference pattern. This can be done  by applying either a dc voltage $V_G^{\mathrm{dc}}$ or a fast time-dependent one $V_G^{\mathrm{ac}}(t)$. Ohmic contacts represented in purple are used for the generation of short single electron pulses by applying a Lorentzian shape \cite{levitov1996,gabelli2013,Dubois13, Safi2022} voltage drive $V_{\mathrm{pulse}}(t)$, and for the measurement of the transmitted dc current $I_\mathrm{out}$ and of its fluctuations $S$. Both the fast time-dependent voltage $V_G^{\mathrm{ac}}(t)$ and the periodic train of single electron pulses $V_{\mathrm{pulse}}(t)$ are generated by an arbitrary wave generator (AWG) with a time resolution of \SI{15.6}{ps} and at a frequency $f=$~\SI{1}{\giga\hertz}. As shown on Fig.  \ref{fig1}C, we label $\tau_e$ the temporal width of the single electron pulse $V_{\mathrm{pulse}}(t)$ (with $\tau_e \approx$ a few tens of picoseconds) and $\tau_s=\frac{1}{2f}=500$ ps that of the square voltage $V_G^{\mathrm{ac}}(t)$. The time delay $t_0$ governs the sampling of $V_G^{\mathrm{ac}}(t)$ by $V_\mathrm{pulse}(t)$.

Before studying single electron interferometry, we first characterize the emitted single electron excitations. $V_{\mathrm{pulse}}(t)$ is a periodic train  of Lorentzian voltage pulses, parameterized by the charge carried by each pulse $qe =\int_{0}^{2\tau_s}{\frac{e^2}{h}V_\mathrm{pulse}(t)\mathrm{d}t}$ and by the temporal width of the pulses $\tau_e$: $V_{\mathrm{pulse}}(t)= \sum_n \frac{q h \tau_e}{\pi e}\frac{1}{(t-n/f)^2+\tau_e^2}$. $q$ and $\tau_e$ are calibrated by measuring the noise $S$ generated by the partitioning of the current $I_{\mathrm{in}}(t)= V_{\mathrm{pulse}}(t)\times e^2/h$ at QPC A (while QPC B is fully open, $T_2=1$, see Fig. \ref{fig1}D). $q$ is extracted from the measurement of $S$ as a function of both the amplitude of the excitation drive generated by the arbitrary waveform generator and the temporal width $\tau_e$ of the pulses \cite{Supp}.  $\tau_e$ is calibrated by performing Hong-Ou-Mandel interferometry \cite{Hong87,Bocquillon13,Dubois13} at QPC A. Two identical trains of single electron pulses are generated at both inputs of QPC A with a tunable time delay $\tau$ between the two inputs (see Fig. \ref{fig1}D). For large time delays ($|\tau| \gg \tau_e$), single electron excitations generated at the two inputs are independently partitioned, and the noise equals the classical random partition noise $\Delta S_\mathrm{class}$ (where $\Delta S$ refers to the excess noise with respect to equilibrium). In contrast, for short time delays $|\tau| \leq \tau_e$, fermionic antibunching at QPC A suppresses the output noise and $\Delta S$ decreases close to 0. 

Fig. \ref{fig1}E presents the measurement of the normalized noise $\Delta S (\tau)/\Delta S_\mathrm{class}$ for two different widths of the pulses generated by the AWG: $\tau_\mathrm{AWG}=$\SI{15.6}{\pico\second} (blue points) and $\tau_\mathrm{AWG}=$~\SI{46.9}{\pico\second} (orange points). The measured HOM dips are then fitted with a Lorentzian shape, providing an \textit{in-situ} measurement of the width of the emitted pulses: $\tau_e=35 \pm 2$~\SI{}{\pico\second} (blue points) and $\tau_e=63 \pm 3$~\SI{}{\pico\second} (orange points). The increase of the measured width $\tau_e$ compared to $\tau_\mathrm{AWG}$ can be explained by the dispersion of the applied voltage pulse when propagating from the AWG to the sample. Fig. \ref{fig1}F gathers our measurements of the width $\tau_e$ for generated pulses of increasing width $\tau_\mathrm{AWG}$ (the red points correspond to $q=1$ pulses and the purple points to $q=2$). We observe that the widening of the pulses is well captured by an offset of \SI{20}{\pico\second} of $\tau_e$ with respect to $\tau_\mathrm{AWG}$.

\section*{Single electron interferometry}

\begin{figure}[h]
    \includegraphics[width = 0.5\textwidth]{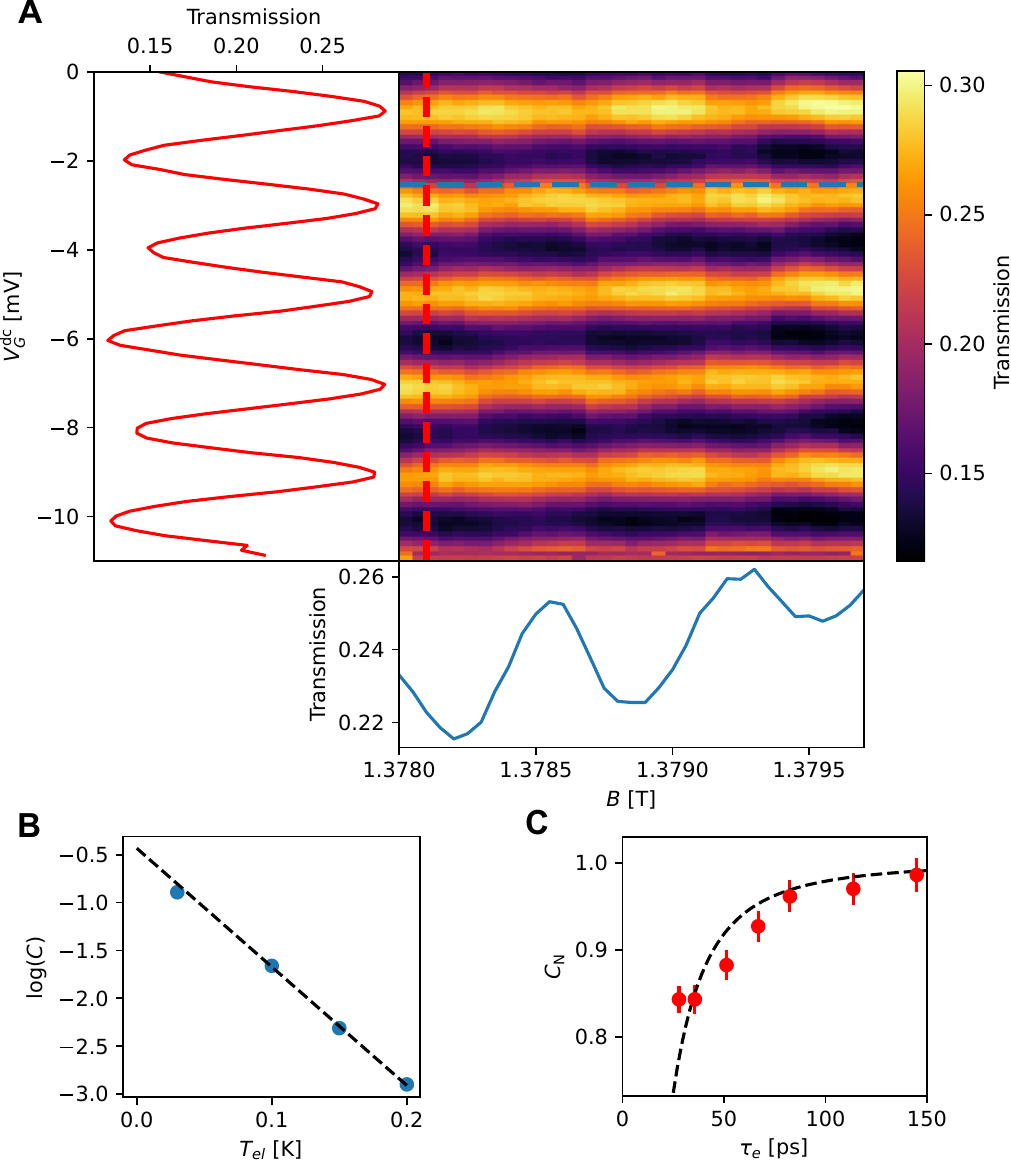}
    \caption{\textbf{Fabry-Perot interferometry with single electrons.} (\textbf{A}) Transmission of the Fabry-Perot interferometer as a function of the magnetic field $B$ and gate voltage $V_{G}^\mathrm{dc}$ showing a periodic behavior. The two insets show cuts along the dotted lines drawn on the central plot as a function of the field (blue) and the dc gate voltage (red). (\textbf{B}) Temperature dependence of the contrast of FP oscillations. (\textbf{C}) Evolution of the normalized contrast of oscillations $C_\mathrm{N}$ as a function of the width $\tau_e$ of single electron Lorentzian pulses. The dashed line represents the overlap of two-Lorentzian wavefunctions $C_N(\tau_L/\tau_e)$ as a function of $\tau_e$ with $\tau_L=$~\SI{30}{\pico\second}.}
    \label{fig2}
\end{figure}

We now move to the measurement of single electron interferences through the FPI by partitioning the outer channel at both QPC A and QPC B. Fig. \ref{fig2}A represents the two-dimensional color plot of the transmission probability $T(V_G^\mathrm{dc},B)$ of single electron excitations through the cavity as a function of the dc plunger gate voltage $V_G^\mathrm{dc}$ and magnetic field $B$. $T$ is extracted from the measurement of the dc current $I_\mathrm{out}$, $T=I_\mathrm{out}/I_\mathrm{in}$, where $I_\mathrm{in}$ is the dc contribution of $I_\mathrm{in}(t)$, $I_\mathrm{in}=q e f$ . We observe large oscillations of $T$ when varying $V_G^\mathrm{dc}$, with a period $\Delta V_G^\mathrm{dc}=$~\SI{2.05}{\milli\volt} and a peak-to-peak amplitude $\Delta T=0.15$ corresponding to an interference contrast $C=\Delta T/(2\left<T\right>)=0.35$. The oscillatory pattern is well reproduced by a sinusoidal fit \cite{Supp}. This shows that $T(V_G^\mathrm{dc},B)$ results from the interference of a single electron with itself after performing one round-trip inside the FPI. We also observe oscillations of $T$ when varying $B$ with a period $\Delta B=0.65 \pm 0.05$~\SI{}{\milli\tesla}. This periodicity corresponds to a variation of $2\pi$ of the Aharanov-Bohm (AB) phase acquired on an area $h/(e\Delta B)= 6.4 \pm 0.5$~\SI{}{\micro\meter\squared}, which matches well the area $A$ of the FPI. However, the amplitude of the $B$ oscillations are roughly five times smaller than the measured oscillations as a function of $V_G^\mathrm{dc}$. 

As discussed in Ref. \cite{Halperin11}, this can be explained by interaction effects within the Fabry-Perot cavity. For strong interactions, corresponding to the Coulomb dominated regime, varying the magnetic field leads to a variation of the interferometer area $A$. This change in $A$ maintains the AB phase constant and suppresses completely the variation of $T$ with  $B$ when the interfering channel is the outer one \cite{Ofek10}. As observed in \cite{Sivan16}, our device is in an intermediate regime, where AB oscillations as a function of $B$ can still be observed yet with a smaller amplitude compared to plunger gate voltage oscillations. In the following, we focus on the measurement of the interference pattern as a function of $V_G^\mathrm{dc}$.

In order to further characterize the FPI, we measure in Fig. \ref{fig2}B the evolution of the interference contrast $C$ as a function of the temperature $T_\mathrm{el}$. The decay of the contrast is well reproduced by an exponential decay with a characteristic temperature scale $T_\mathrm{el}^0=$~\SI{81}{\milli\kelvin}. Considering a thermal averaging of the interference pattern \cite{Chamon97}, $T_\mathrm{el}^0= \hbar/(\pi k_B \tau_L)$ is related to the time $\tau_L=L/v$ (where $v$ is the electron velocity) it takes for an electron to make one round-trip in the cavity. This allows us to estimate $\tau_L=$~\SI{30}{\pico\second} corresponding to a velocity $v=3.8\times 10^5$~m.s$^{-1}$. 


For short single electron pulses, the finite travel time inside the cavity $\tau_L$ leads to a reduced overlap of the electronic wavefunction at the interferometer's output, leading to a reduced interference contrast. For Lorentzian wavepackets of wavefunction $\varphi_{\tau_e}(t)=\frac{\sqrt{\tau_e/\pi}}{t-i \tau_e}$, the overlap is given by $C_N(\tau_L/\tau_e)= \Re [ \int dt  \varphi_{\tau_e}(t) \varphi^*_{\tau_e}(t+\tau_L) ] = \frac{1}{1+ (\tau_L/2\tau_e)^2}$.  $C_N(\tau_L/\tau_e)$ is also the interference contrast normalized by its value for $\tau_L=0$,  $C_N(\tau_L/\tau_e)=C(\tau_L/\tau_e)/C(\tau_L/\tau_e=0)$. Fig. \ref{fig2}C represents the evolution of $C_N(\tau_L/\tau_e)$ as a function of $\tau_e$. In the limit $2\tau_e > \tau_L$, a small reduction of $C_N$ is expected. We indeed observe such a small reduction that can be accounted for by the above expression of $C_N(\tau_L/\tau_e)$ using $\tau_L=$~\SI{30}{\pico\second} (see also \cite{Supp}).

\section*{Single electron interferometry for the detection of fast electric fields}

\begin{figure*}
    \includegraphics[width = \textwidth]{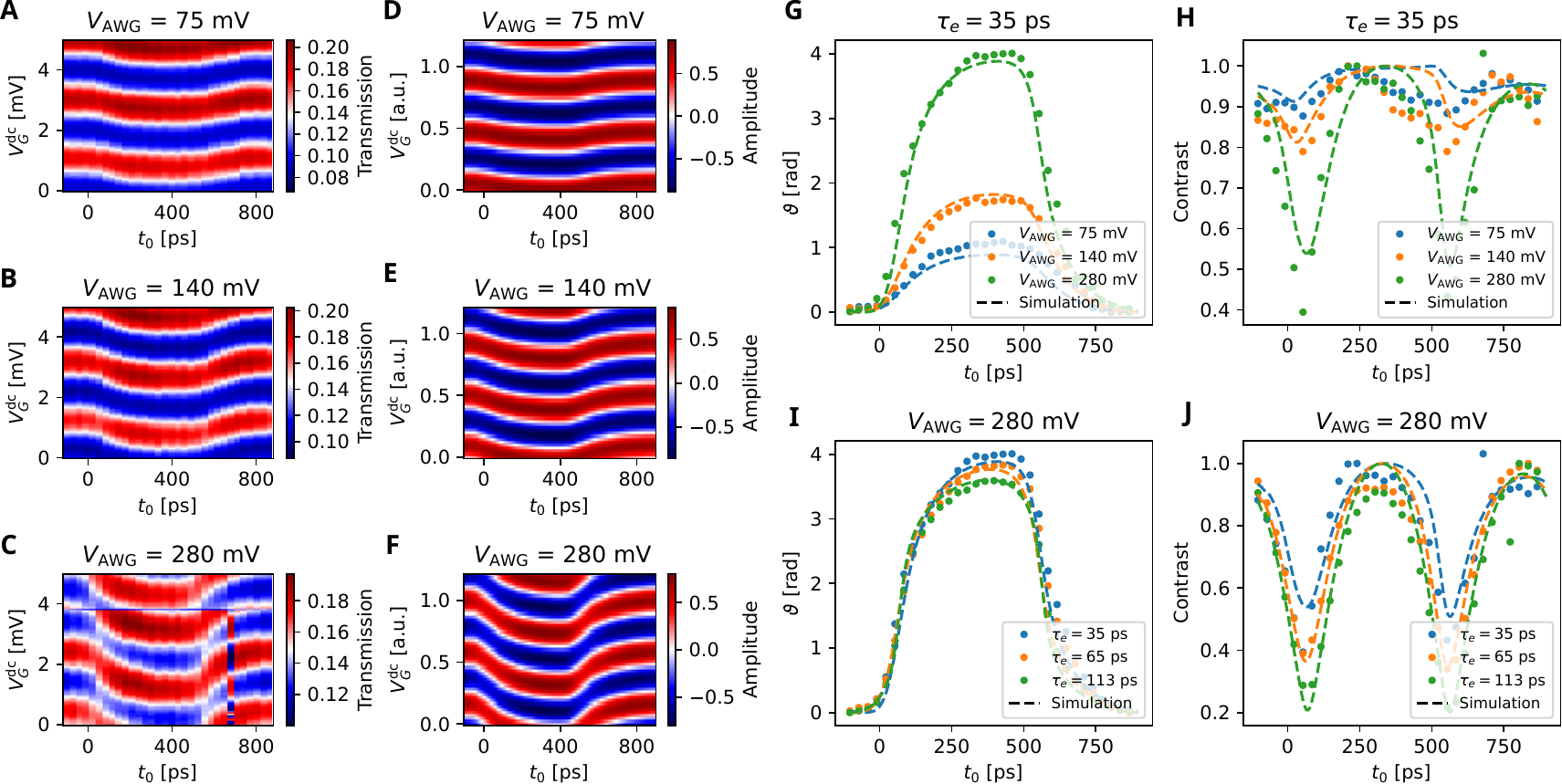}
    \caption{\textbf{Sensing of a time-dependent voltage with single electron interferometry.} (\textbf{A})-(\textbf{C}) Transmission measured at the output of the FPI as a function of the delay $t_0$ between the single electron pulses $V_\mathrm{pulse}(t)$  and the square excitation $V_G^\mathrm{ac}(t)$. The three maps show data for Lorentzian pulses of width \SI{35}{\pico\second} with varying amplitude of the square excitation $V_\mathrm{AWG}=75$, $140$ and \SI{280}{\milli\volt}. (\textbf{D})-(\textbf{F}) Simulations performed using the same parameters as in A-C and rise time of the square \SI{140}{\pico\second}. (\textbf{G}) Evolution of the phase $\vartheta$ of the oscillations as a function of the delay $t_0$ for a fixed width $\tau_e=$~\SI{35}{\pico\second} of the single electron pulse varying the amplitude $V_\mathrm{AWG}$ of the square drive. (\textbf{H}) Associated contrast $C(t_0)$. (\textbf{I}) Evolution of the phase $\vartheta$ of oscillations as a function of the delay $t_0$ for three widths $\tau_e$ of the single electron pulses. (\textbf{J}) Associated contrasts $C(t_0)$. The full data sets used to obtain these curves can be found in the supplementary section of the paper. The simulation are shown in dashed lines in figures G-H.}
    \label{fig3}
\end{figure*}

We now describe our main realization that consists in the detection of fast electric fields using single electron interferometry. The signal detected is the change in the interference current that results from the time-dependent voltage $V_{G}^\mathrm{ac}(t)$ applied on the plunger gate (see Fig. \ref{fig1}A). We choose a square-shaped voltage of temporal width $\tau_s=$~\SI{500}{\pico\second} and repetition frequency $f=$~\SI{1}{\giga\hertz}. We vary the peak-to-peak amplitude of the generated square from $V_\mathrm{AWG}=$~\SI{75}{\milli\volt} to $V_\mathrm{AWG}=$~\SI{280}{\milli\volt}. $V_\mathrm{AWG}$ is the peak-to-peak amplitude generated at room temperature. After being attenuated at each stage of the fridge, it corresponds to a peak-to-peak amplitude $V_{G}^\mathrm{ac}$ at the level of the plunger gate that varies from \SI{350}{\micro \volt}  (for $V_\mathrm{AWG}=75$ mV) to \SI{1.3}{\milli \volt} (for $V_\mathrm{AWG}=280$ mV). $V_{G}^\mathrm{ac}(t)$ is then detected by measuring the interference pattern $T(V_G^\mathrm{dc},t_0)$ as a function of both the dc plunger gate voltage $V_{G}^\mathrm{dc}$ and the time delay $t_0$ between $V_{G}^\mathrm{ac}(t)$ and $V_{\mathrm{pulse}}(t)$ (see Fig\ref{fig1}c). 

Figs. \ref{fig3}A-C represent the two-dimensional color plot of $T(V_G^\mathrm{dc},t_0)$ for three different amplitudes of the square voltage excitation, $V_\mathrm{AWG}=75$, $140$ and $280$ mV. The temporal width of the emitted single electron pulses is $\tau_e=$~\SI{35}{\pico\second}. The observed effect of $V_{G}^\mathrm{ac}(t)$ on $T(V_{G}^\mathrm{dc},t_0)$ can be easily understood. $V_{G}^\mathrm{ac}(t)$ leads to a phase shift of the interference pattern for $t_0\approx 100$ and $t_0=$~\SI{600}{\pico\second} corresponding to the time where the emission of a single electron is synchronized with the sudden variations of the square voltage excitation. As expected, the measured phase shift increases when the amplitude of the square voltage varies from \SI{75}{\milli\volt} to \SI{280}{\milli\volt}. To extract the temporal variation of $V_{G}^\mathrm{ac}(t)$, we measure for each time delay $t_0$ the complex contrast $C(t_0) e^{i \vartheta(t_0)}$ of the interference pattern from a sinusoidal fit of $T(V_{G}^\mathrm{dc})$ \cite{Supp}. For each amplitude $V_\mathrm{AWG}$, we choose a phase reference $\vartheta(t_0^\mathrm{ref})=0$ taken at the first data point $t_0^\mathrm{ref}=$\SI{-100}{\pico\second}.

Fig. \ref{fig3}G represents our measurement of $\vartheta(t_0)$ for the three amplitudes of the square excitation $V_\mathrm{AWG}$. The shape of the square excitation is well reproduced. As detailed below, for such short electronic wavepackets with $\tau_e=$~\SI{35}{\pico\second}, the temporal resolution of our voltage measurement is mainly limited by the rise time of \SI{140}{\pico\second} of the applied square excitation $V_{G}^\mathrm{ac}(t)$  and not by our experimental detection method. Importantly, we observe that the measured phase shift $\vartheta(t_0)$ scales linearly with the excitation amplitude up to error bars. This implies that our method directly reconstructs $V_{G}^\mathrm{ac}(t)$ from the measurement of $\vartheta(t_0)$, with $V_{G}^\mathrm{ac}(t_0)=\frac{e}{C_G}\frac{\vartheta(t_{0})}{2\pi}$, where $C_G=\SI{0.08}{\femto \farad}$ is deduced from the dc plunger gate voltage periodicity $\Delta V_G^\mathrm{dc}$, $C_G=e/\Delta V_G^\mathrm{dc}$. This linear relation between the interference phase and the detected voltage is important for accurate reconstructions of voltages in a large dynamical range as well as for future applications for quantum signals. Our voltage resolution is $\approx$ \SI{50}{\micro \volt}, taken as three times the error bar of the reconstructed phase signal. Note that the sensitivity could be increased by a factor 10 by increasing the size of the plunger gate accordingly, thereby increasing the gate capacitance and reaching the few \SI{}{\micro \volt} sensitivity. However, this would also slightly decrease the detection time resolution by increasing the coupling time between the gate and the single electronic wavepackets.  

The quantum nature of our detection process is nicely illustrated by Fig. \ref{fig3}H that presents the contrast of the interference  $C(t_0)$ extracted from the sinusoidal fit mentioned above. In order to compare together the different amplitudes of the detected square voltage excitation $V_\mathrm{AWG}$, we plot on Fig. \ref{fig3}H the contrast normalized by the maximal value it reaches when varying $t_0$. For all traces, a clear suppression of the contrast is observed for $t_0 \approx 80$ and $t_0 \approx$~\SI{580}{\pico\second}, corresponding to the times for which $V_{G}^\mathrm{ac}(t_0)$ rises up and falls down. Close to these two values of $t_0$, the different temporal components of the interfering electronic wavepacket (with a characteristic width $\tau_e$) experience different values of the interference phase (corresponding to different values of the square plunger gate voltage). This leads to a reduction of the interference contrast. As observed in Fig. \ref{fig3}H the contrast reduction is more pronounced when  $V_\mathrm{AWG}$ increases, which leads to an increased spreading of the phase acquired by the different components of the electronic wavepacket. This contrast reduction strikingly demonstrates the quantum nature of the detection process: the interference contrast is reduced by the quantum fluctuations of the position within the single electronic wave packet. 

The role of the temporal width of the emitted wavepackets is illustrated on Figs. \ref{fig3}I and J, representing $\vartheta(t_0)$ and $C(t_0)$ for a fixed amplitude $V_\mathrm{AWG}=$~\SI{280}{\milli\volt} and different wave packet widths $\tau_e=35$, $65$ and \SI{113}{\pico\second}. Fig. \ref{fig3}I shows the importance of using short wave packets for a better time resolution of the reconstruction of $V_G^\mathrm{ac}(t)$. As observed on Fig. \ref{fig3}I, the extracted temporal evolution of $\vartheta(t_0)$ is smoothed when increasing $\tau_e$, reducing the amplitude of variation of $\vartheta(t_0)$ and increasing its rise time. As seen on Fig. \ref{fig3}J, increasing the spread of the electronic wave packet also enhances the reduction of the contrast by the quantum fluctuations of the electron position. The contrast dips gets more and more pronounced when increasing $\tau_e$ from \SI{35}{\pico\second} to \SI{113}{\pico\second}.

Our measurements can be well reproduced using a simple model of single electron interference introduced in Ref.\cite{Souquet24} (see also \cite{Supp}). We compute the complex contrast of interference $C(t_0)e^{i\vartheta(t_0)}$ in the presence of the time-dependent modulation $V_G^\mathrm{ac}(t)$ normalized by its value for $V_G^\mathrm{ac}(t)=0$:

\begin{eqnarray}
C(t_0)e^{i\vartheta(t_0)} & =& \frac{\int \mathrm{d}t e^{i 2\pi \frac{C_G}{e}V_G^\mathrm{ac}(t)} \varphi_{\tau_e}^*(t_0-t) \varphi_{\tau_e}(t_0-t-\tau_L) }{ \int \mathrm{d}t \varphi^*_{\tau_e}(t) \varphi_{\tau_e}(t-\tau_L)}. \nonumber\\ \label{eq1}
\end{eqnarray}

The model (in dashed lines on Fig. \ref{fig3}G-J) reproduces very well all the experimental observations, such as the evolution of the phase $\vartheta(t_0)$, its smoothing when increasing the width of the emitted wavepackets $\tau_e$, as well as the decrease of the contrast $C(t_0)$ due to the quantum fluctuations of the electron position. The nice agreement between data and model demonstrates our ability to probe time-dependent voltages by exploiting the quantum phase of a single electron wavefunction.

\section*{Conclusion}
We have demonstrated that single electron quantum states could be used as a fast and sensitive probe of time-dependent voltages. By measuring the phase $\vartheta(t_0)$ of a single electron interference pattern in a FPI, we reconstruct a time-dependent voltage $V_G^\mathrm{ac}(t_0)$ applied to a metallic gate coupled to one arm of the interferometer.  We reach a time resolution of a few tens of picoseconds, limited by the temporal width $\tau_e$ of the emitted  wavepackets, and a voltage resolution of a few tens of microvolts. The voltage resolution could be improved by increasing the size of the metallic gate that couples the probed electromagnetic field to the interferometer, or by increasing the emission frequency of single electrons, ultimately reaching the \SI{}{\micro \volt} resolution. The measurement of the contrast $C(t_0)$ demonstrates the quantum nature of our detection method. We observe a sharp decrease of $C(t_0)$ close to fast variations of $\vartheta(t_0)$ caused by the quantum fluctuations of the electron's position. 

The results presented here concern the detection of classical voltages. Our method can be extended to exotic quantum states of the electromagnetic field, such as Fock or squeezed states generated on-chip\cite{gasse2013,Bartolomei23}. In the latter case, measuring the enhancement and reduction of the  contrast $C$ when varying $t_0$ would directly reflect the enhancement or reduction of the phase fluctuations associated to the squeezing of the electromagnetic field. Finally, the use of the phase of a single electron wavefunction as the basic element of a quantum detection scheme opens the way for completely new detection methods. For example, one could engineer the electron phase using chirping methods (a quadratic temporal variation  $\propto \kappa t^2$  of the phase of the electronic wavefunction as discussed in \cite{Keeling08} for example).  Such chirped wavepackets exhibit a time dependent spectral content. Fed by such wavepackets, the interferometer will exploit the beating between the spectral contents at two different times separated by $\tau_L$, thereby downconverting the frequency of the probed external phase by $\kappa\tau_L$, corresponding to the \SI{}{\tera\hertz} band for an energy ramp $\kappa$ of the order of \SI{1}{\milli \volt} on a few tens of picoseconds. This opens a new  and challenging road for the engineering of single electron states for quantum sensing applications. 

\appendix 

\section*{Methods}

\subsection*{Current noise measurements}

The current noise at the output of QPC A in the HOM configuration is converted to a voltage noise via the quantum Hall edge channel resistance $R_\nu = h/\nu e^2$ between the output ohmic contact and the ground.  In order to move the noise measurement frequency in the MHz range to avoid parasitic noise contribution at low frequency, the output ohmic contact is also connected to the ground via an LC tank circuit of resonance frequency $f_0 = 1.1$ MHz. The tank circuit is followed by an homemade cryogenic amplifier and a room temperature amplifier. A vector signal analyzer measures the autocorrelation of the output voltage noise in a \SI{100}{\kilo\hertz} bandwidth centered on $f_0$. The current noise measurements are calibrated by measuring the thermal noise of the output resistance $R_\nu$ as a function of the temperature.

\subsection*{Average current measurements}

The dc contribution of the output current $I_{\mathrm{out}}$ generated by the periodic train of single electron pulses $V_{\mathrm{pulse}}(t)$ is measured by a lock-in amplifier by applying a square modulation to $V_{\mathrm{pulse}}(t)$. The modulation is performed at \SI{1}{\mega\hertz}, thus averaging over many pulses generated with a \SI{1}{\giga\hertz} frequency, alternating sign at \SI{1}{\mega\hertz}. $I_{\mathrm{out}}$ is converted into a voltage signal on the output impedance of the sample $Z$ that consists in the Hall resistance $R_\nu$ in parallel with the LC tank circuit described above. It is then amplified with total gain $G$ by a homemade cryogenic amplifier followed by a commercial room temperature amplifier. The current $I_{\mathrm{out}}$ as well as the charge $qe$ carried by each pulse are calibrated following a procedure described below (see also \cite{Supp}). 

\subsection*{Calibration of the single electron pulses}

The charge $qe$ carried by each pulse is determined by measuring both the calibrated excess current noise $\Delta S$ generated by the partitioning of the single electron pulses by QPC A and the uncalibrated dc input current  $I_{\mathrm{in}}$. We plot on the same graph (see \cite{Supp}) the noise measurements $\frac{\Delta S}{T_1(1-T_1)}$ obtained for different amplitudes and different widths $\tau_e$ of the generated voltage pulses $V_{\mathrm{pulse}}(t)$. These noise measurements are plotted as a function of our measurements of the uncalibrated amplified dc input current $G |Z|  I_{\mathrm{in}}$. Remarkably, all data points fall on a linear slope, reflecting that the noise is proportional to the input current: $\frac{\Delta S}{T_1(1-T_1)}=2 e I_{\mathrm{in}}= 2 e^2 f q$,  where $f=$~\SI{1}{\giga\hertz} is the repetition frequency. We can thus calibrate the lever arm $\alpha$ relating our measurement of the input current to the charge $q$, $\alpha= G|Z|I_\mathrm{in}/q$. By choosing $\alpha=1.81\cdot 10^{-4}$ V, we impose that our data $\frac{\Delta S}{T_1(1-T_1)}$ have the expected slope $2e^2 f$ when plotted as a function of $q= G|Z|I_\mathrm{in}/\alpha$. This provides both a calibration of the charge per pulse $q$ and of the input current $I_{\mathrm{in}}$. We can check the soundness of our calibration procedure by plotting also the noise $\frac{\Delta S}{T_1(1-T_1)}$ generated by a dc voltage bias $V_{\mathrm{dc}}$, with $q_{\mathrm{dc}}=e V_{\mathrm{dc}}/(hf)$.  As can be seen in \cite{Supp}, all our measurements fall nicely on the expected slope for shot noise $\frac{\Delta S}{T_1(1-T_1)}= 2 e^2f q$.

\subsection*{Extraction of the phase and contrast of the single electron interferometric signal}

In order to extract the phase and contrast of the single electron interferometric signal, we perform cuts on the two-dimensional maps $T(V_G^\mathrm{dc},t_0)$  at fixed $t_0$. These cuts show an oscillating signal as a function of $V_G^\mathrm{dc}$ (see Fig. \ref{fig2}A) which is fitted using a cos function of the form $C(t_0) \cos{(V_G^\mathrm{dc}/V_0+\vartheta(t_0))}+b$. The fit parameter $\vartheta(t_0)$ is then used to plot Fig. \ref{fig3}G and Fig. \ref{fig3}I. The contrast plotted in  Fig. \ref{fig3}H and Fig. \ref{fig3}J is then calculated as the ratio $C(t_0)/\mathrm{max}\big(C(t_0)\big)$. From these fits we observe that there is no second harmonic contribution to the signal and that a simple sinusoidal oscillation describes our experimental data perfectly, justifying the use of a model where a single round-trip inside the FP cavity is taken into account. 

\section*{Acknowledgements}
The authors thank H. Souquet-Basiège, B. Roussel, V. Kashcheyevs and C. Bäuerle for fruitful discussions during the preparation of this manuscript.

\section*{Funding}
This project received funding from the Agence Nationale de la Recherche under the France 2030 programme, reference ANR-22-PETQ-0012, and from the  ANR grant ``QuSig4QuSense'' (ANR-21-CE47-0012). This work is supported by the French RENATECH network. The authors have applied a CC-BY public copyright licence to any Author Accepted Manuscript (AAM) version arising from this submission.

\section*{Author contributions}
YJ fabricated the sample on GaAs/AlGaAs heterostructures grown by AC and UG. YJ designed and fabricated the low-frequency cryogenic amplifiers used for noise measurements. HB, EF and MR conducted the measurements. HB, EF, MR, EB, JMB,  GM and GF participated to the data analysis and the writing of the manuscript with inputs from GR, IS, PD, YJ and UG. Theory was done by GR, IS and PD.  GF and GM supervised the project.

\section*{Competing interests}

The authors declare that they have no competing interests.

\appendix 
\section{Figure of merit of the detection}

\begin{figure}[b]
    \centering
    \includegraphics[width = 0.6\textwidth]{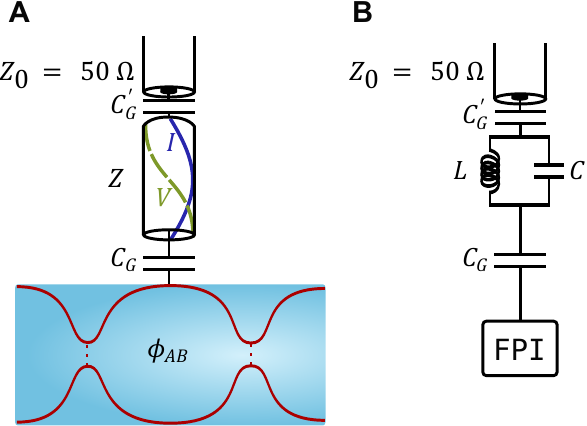}
    \caption{\textbf{Estimating the detector figure of merit:} \textbf{(A)} We consider an rf resonator containing one photon. The ground state is associated to a given current $I$ and voltage $V$ that couples to the FPI through the capacitance $C_G$. \textbf{(B)} This configuration can be seen as an LC resonator of capacitance $C$ and inductance $L$ capacitively coupled to the FP.}
    \label{figS12}
\end{figure}

Let us consider zero-point fluctuations in circuit QED like systems (see Fig.~\ref{figS12}.A). The charge operator can be written as
\begin{equation}
    \hat{Q} = -iQ_\mathrm{ZPF}(\hat{a}-\hat{a}^{\dagger}),
\end{equation}
where $Q_\mathrm{ZPF}=\sqrt{\hbar/2Z}$, with $Z$ the characteristic impedance of the line (see Fig.~\ref{figS12}.A), typically equal to \SI{50}{\ohm}. Taking the mean value of the square of the charge leads to $\braket{\hat{Q}^2}_{\ket{N}}\simeq Q_\mathrm{ZPF}^2(2N+1)$ where $N$ is the mean number of photons in our circuit. The charge $Q$ is linked to the voltage $V$ through the capacitance $C$ such that $V=Q/C$, which allows us to write the voltage fluctuations in the $N$ photon Fock state $\ket{N}$ of the LC-resonator:
\begin{equation}
    \langle V^2\rangle_{\ket{N}}=\left(\frac{Q_\mathrm{ZPF}}{C}\right)^2(2N+1).
\end{equation}
The approximate voltage increase $\Delta V_\mathrm{1ph}$ associated to the presence of a single photon $(N=1)$ in the circuit is $\Delta V_\mathrm{1ph}\simeq\sqrt{3}\frac{Q_\mathrm{ZPF}}{C}$.  Rewriting $Q_\mathrm{ZPF}$ as $e\sqrt{R_K/4\pi Z}$ where $R_K$ is the quantum of resistance $h/e^2$ leads to $\Delta V_\mathrm{1ph}\simeq \sqrt{\frac{\hbar}{Z}}\frac{1}{C}$. The characteristic impedance of the line can be written in terms of the line inductance and capacitance as $Z=\sqrt{L/C}$, and so can the frequency $2\pi f=1/\sqrt{LC}$. Combining those quantities we have
\begin{equation}
    e\Delta V_\mathrm{1ph}=\frac{e^2}{C}\sqrt{\frac{\hbar}{Ze^2}}=\frac{e^2}{C}\sqrt{\frac{R_K}{2\pi Z}}
	=\frac{e^2}{C}\frac{1}{\sqrt{2\pi
	z}}=\frac{e^2}{C}\frac{hf}{\sqrt{2\pi z}}
	\frac{\sqrt{LC}}{\hbar}=hf\,\sqrt{2\pi z}.,
\end{equation}
where $z$ is the ratio of the characteristic impedance of the line and the quantum of resistance $Z/R_K$.

Assuming a \SI{50}{\ohm} characteristic impedance of the line we have in our experiment $z=2\times 10^{-3}$. Applying this result to our measurement at $f=$~\SI{10}{\giga\hertz} we obtain an equivalent voltage associated to the presence of a single photon $\Delta V_\mathrm{1ph}\simeq$~\SI{4}{\micro\volt}. Therefore, with an estimated experimental voltage resolution of \SI{50}{\micro\volt} our current apparatus has an equivalent detection resolution of about 10 photons. This number could be greatly improved by modifying the geometric parameters of the driving gate.

\section{Presentation of the setup and modelization}
\label{sec/setup}

	
\begin{figure}[h!]
\centering
\includegraphics[scale=0.4]{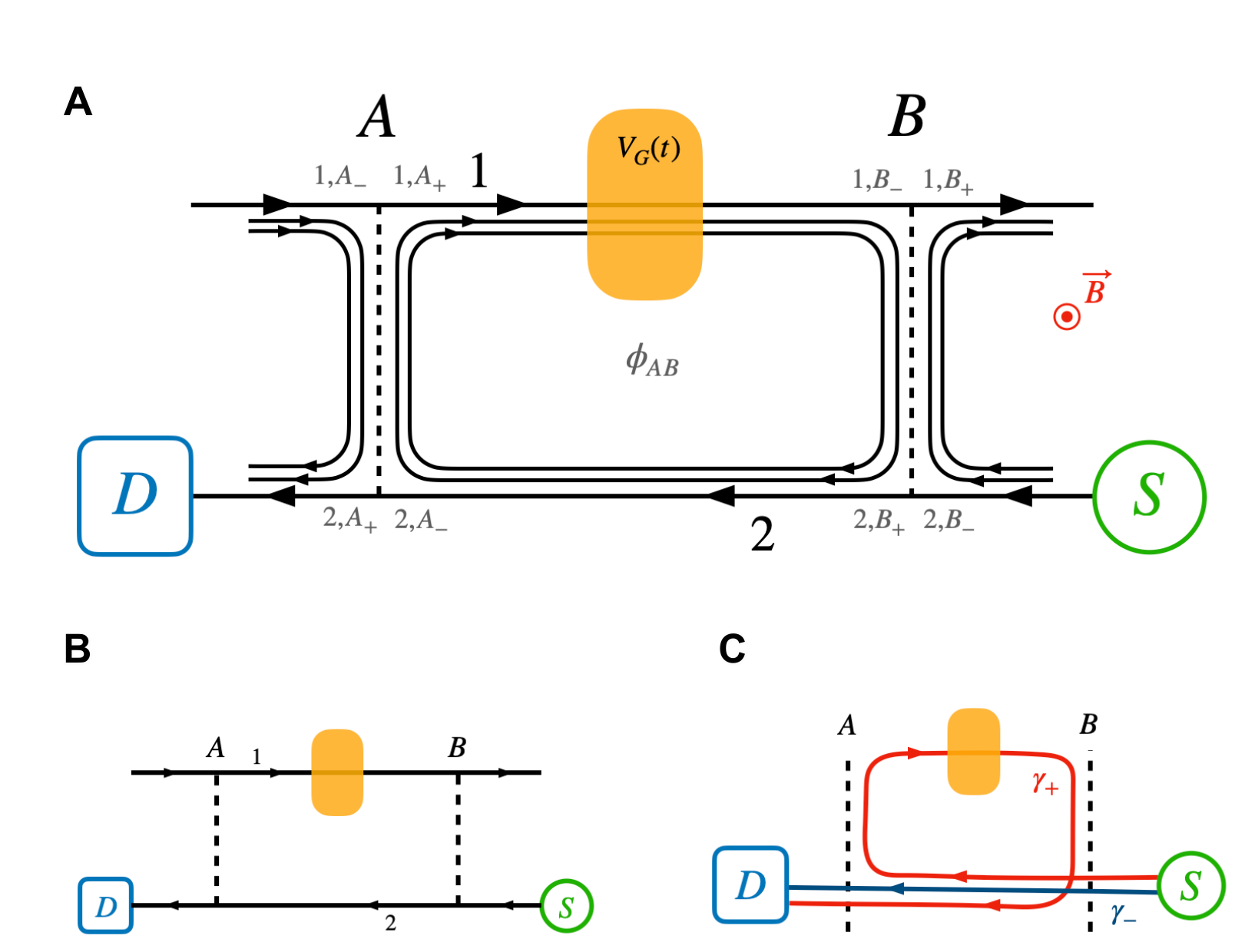}
\caption{\label{fig/FP} \textbf{(A)} Full schematic of the Fabry-Perot interferometer at $\nu=3$. The QPCs $A$ and $B$ operate
in the weak-backscattering (WB) regime where the transport channel is the outer one. The orange box in the upper branch is the plunger gate $V_G(t)$ we want to probe by measuring the interference contribution to the average dc current  at the detector $D$.  \textbf{(B)}  The Fabry-Perot interferometer in the weak backs-cattering regime with only charge transporting channel represented.  \textbf{(C)}  Visualization of the interference paths $\gamma_{\pm}$  from the source $S$ to the detection $D$ in this configuration.
}
\end{figure}
	
\subsection{Fabry-Perot configuration}
	
The Fabry-Perot (FP) interferometer set-up is presented in Fig.~\ref{fig/FP}A  where single electrons are injected from the source $S$. The two QPCs behave as ideal electronic beam splitters. A perpendicular magnetic field is applied, generating an Aharonov-Bohm phase
$\phi_\mathrm{AB}$ which, in full generality, is related to the magnetic flux $\Phi_B$ enclosed between the arms $1$ and $2$ as well as on the dc-component of the plunger gate voltage $V_G^{\text{dc}}$ (see Ref.~\cite{Halperin-2011-1} and a discussion specific to the device
considered here in Sec.~\ref{sec/FP/coupled-branches}). The classical time-dependent voltage $V_G(t)$ in the orange box is probed through the measurement of the average current at the detector $D$ which reads
\begin{equation}
	i_D(t)=\langle \hat{\psi}^\dagger_D(t)
	\hat{\psi}_D(t) \rangle_S =-e \bigg( I_0 (t)+
	2\,\mathrm{Re}\big[\me^{-\mi\phi_{AB}} I_+(t)\big] \bigg)
	\label{interf}
\end{equation}
with $ \hat{\psi}_D(t)$ being the fermionic operator at the detector at time $t$ and $I_+(t)$ being the interference contribution to the average electrical current. Here $\langle \ldots \rangle_S$ denotes an average taken over the many body state corresponding to the source $S$ switched on.
	
Since the interferometer is operating in the weak-backscattering (WB) limit (i.e. $R_{\alpha}\ll 1$) we expect the dominant paths to the interference to be the ones in Fig.~\ref{fig/FP}C: the single electron can either go straight through the two quantum point contacts (QPCs), reaching the detector (blue path $\gamma_-$), or turn around the inner cavity making one lap and a half (red path $\gamma_+$).
	
\subsection{HOM configuration}

\begin{figure}[h!]
	\centering
	\includegraphics[scale=0.4]{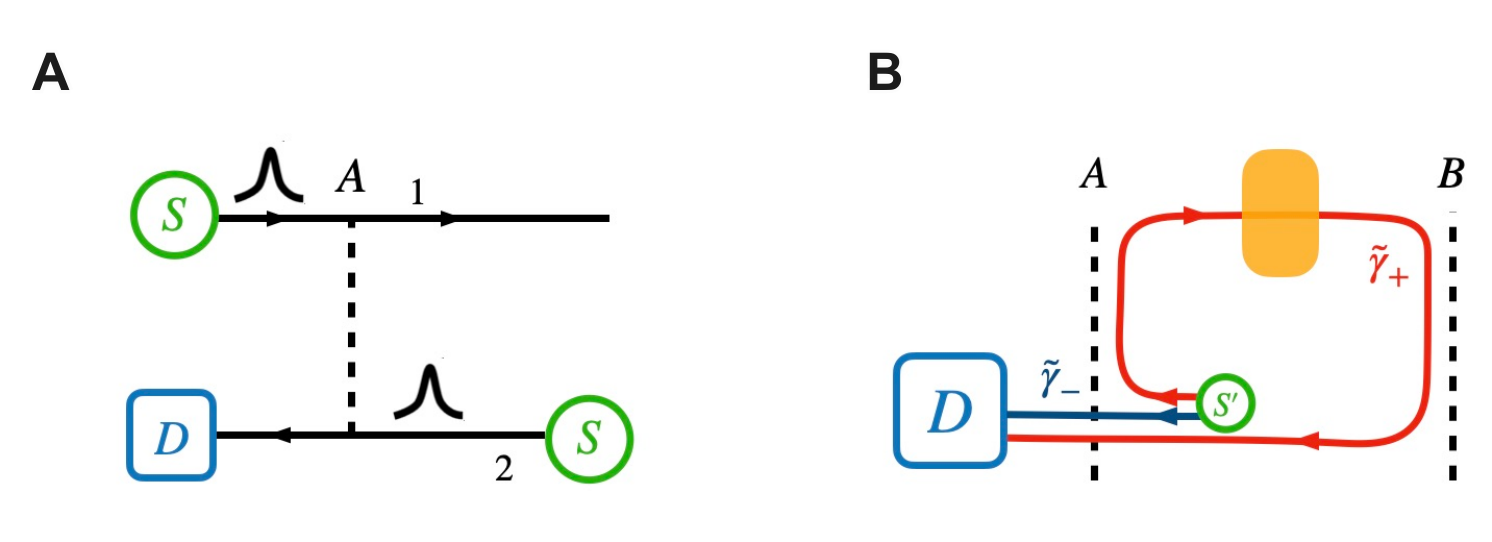}
	\caption{\label{fig:FP_2} \textbf{(A)} The characterization of Lorentzian pulses is done experimentally at QPC $A$ by HOM interferometry. \textbf{(B)} Visualization of the new interference paths $\tilde{\gamma}_\pm$ assuming single electron excitations have been injected right before QPC $A$.}
\end{figure}

The device is also calibrated in a configuration based on Hong-Ou-Mandel (HOM) interferometry as depicted on Fig.~\ref{fig:FP_2}A. In this mode, QPC B is fully open and HOM calibration is performed at QPC A and used to identify Lorentzian voltage pulses there. An important point is that in the calibration modes, the inner channels appearing on Fig.~\ref{fig/FP}A are no longer closed. 
	
This implies that, in this mode, all the channels of branch $2$ are excited symmetrically by $S$. Since $S$ consists of exciting the charge mode of branch $2$, it means that all edge channels of branch $2$ carry the same coherent state of edge-magnetoplasmons (EMPs). In the HOM configuration (see fig.1D-F of main text), HOM interferometry is performed on the outer edge channel and therefore it can be viewed as a characterization of the EMP coherent state within each of these edge channels.

It is important to notice that, in this operation mode, the electromagnetic environment of the outer edge channel of branch $2$ is not the same than in the FP configuration since the geometry of all the other channels is closed.

\subsection{Modeling strategies}

A first difficulty comes from the fact that the calibration mode does not characterize the excitations arriving at QPC A in the FP configuration. In principle, since we are dealing with coherent EMP states, it is possible to infer the coherent state of EMPs injected by $S$ into branch $2$ and then to discuss the FP configuration.
	
The method used to discuss the FP configuration consists of back-propagating the fermionic field $\hat{\psi}_D(t)$ along the two relevant interfering paths thereby formulating it in a way similar to Mach-Zehnder (MZ) interferometry \cite{Souquet-Basiege-2024-1}. The averages are then computed using the state emitted by $S$. 

However, the analogy is not straightforward due to the coupling of the outer channels of the two branches via
the closed edge channels. In order to understand the system, we will discuss increasingly complex models of the FP interferometer that incorporate more and more features that have just been discussed.
	
First of all, in Sec.~\ref{sec/FP/no-interactions}, we neglect Coulomb interactions and use time dependent single particle scattering to obtain the interference contribution to the dc average current. This simple model, despite its limitations, clearly depicts the analogy with MZ interferometry and enables to see the role of the various time of flights in the system. We derive the contrast of the interferences on the dc average current in terms of the overlap between a Leviton wavepacket delayed by the full time of flight around the Fabry-Perot loop and itself and a filter applied to the time dependent phase imprinted on the electrons by the time dependent plunger gate voltage $V_G(t)$.
	
Then, Coulomb interactions are introduced in two steps: first, in Sec.~\ref{sec/FP/uncoupled-branches}, we assume that branches
$1$ and $2$ of the FP interferometer (see Fig.~\ref{fig:FP_2}B) are not coupled via Coulomb interactions. This accounts for Coulomb interactions within each of them as well as capacitive couplings to environmental degrees of freedom, each branch having its own environment. However, as shown on Fig.~\ref{fig/FP}A, this does not account for the presence of closed edge channels when the device is operated in the FP configuration. Nevertheless, when injecting electronic excitations with energies much below any resonance frequency associated with these closed edge channels, we can expect the uncoupled branches model to be valid. The main result of this analysis is that, at very low temperatures, and provided that the time dependent voltage $V_G(t)$ is slow compared to the characteristic times associated with electronic decoherence, the relative contrast between the interference signal at non zero $V_G(t)$ and at $V_G(t)=0$ can be computed using free electron expressions derived in Sec.~\ref{sec/FP/no-interactions}. 

In Section \ref{sec/FP/coupled-branches}, we discuss the case where the two branches are coupled via the closed edge channels and
argue qualitatively that the effect related to the closed edge channels may be weak in the domain of operation of the experiment. 

Finally, in Sec.~\ref{sec/filter-analysis}, we study the relation between the relative contrast and the interference signal at non zero $V_G(t)$ and at $V_G(t)=0$ and the phase imprinted on the electrons by the top gate.

\subsection{Main results}
\label{sec/setup/main-results}

Let us now summarize the main results of the theoretical modelization details in the next sections. The quantity which is measured is the dc average current which depends on the Aharonov-Bohm phase $\phi_{\text{AB}}$ in a $2\pi$-periodic way
\begin{equation}
\langle i_D^{(\text{dc})}\rangle_S\simeq
	-e \left( I_0^{(\text{dc})}+
	\me^{-\mi\phi_{AB}} I_+^{(\text{dc})}+
	\me^{\mi\phi_{\text{AB}}}I_-^{(\text{dc})}\right)
    \label{eq/interf-dc}
\end{equation}
where we only have retained the first harmonics in $\phi_{\text{AB}}$ in accordance with experimental data. Note that the interference
contribution $I_+^{(\text{dc})}=\left(I_-^{(\text{dc})}\right)^*$ depends on the wavepackets injected by $S$, that is their duration $\tau_e$ as well as their injection time $t_0$ and of course, of the time dependent top gate voltage $V_G(t)$.

We define the relative interference contrast  as the ratio of $I_+^{(\text{dc})}$ for a given $V_G(t)$ to the same quantity for $V_G(t)=0$. It is a complex number that depends on experimentally controlled parameters $t_0$, $\tau_e$ as well as $V_G(t)$. Our theory provides an explicit prediction for this relative contrast valid even in the presence of decoherence effects, provided they manifest themselves only at frequencies much higher than the ones involved in $V_G(t)$:
\begin{equation}
	\label{eq/main-results/1}
	C(t_0)\,\me^{\mi\vartheta(t_0)}=\int_{\mathbb{R}}
	\me^{\mi\phi_1(V_G,\tau)}f_{\text{FP}}(t_0-\tau)
	\md\tau
\end{equation}
where $\me^{\mi\phi_1(V_G,\tau)}$ is the phase imprinted on the electrons by the top gate voltage. This expression for the contrast is the general form of eq.1 of the main text. For a short top gate located in the middle of branch $1$, and assuming that $V_G(t)$ does not vary during the characteristic time with the plunger gate capacitively coupled to the edge channel, it can be expressed as
\begin{equation}
	\label{eq/main-results/2}
	\me^{\mi\phi_1(V_G,t)}=\me^{2\pi \mi
	C_GV_G\left(t-\frac{\tau_1}{2}\right)/e}
\end{equation}
in which $C_G$ denotes the electrochemical capacitance associated with the capacitor built from the top gate and the quantum Hall edge channel beneath it. Assuming we use Leviton electronic excitations of wavefunction\footnote{Note that here
and in the rest of this Supplementary, $\varphi_{\tau_e}(t)$ has dimension $[\mathrm{T}]^{-1/2}$.}
\begin{equation}
	\label{eq/Leviton-wavefunction}
	\varphi_{\tau_e}(t)=\sqrt{\frac{\tau_e}{\pi
}}\,\frac{1}{t-\mi\tau_e}\,,
\end{equation}
the filter function $f_{\text{FP}}(\tau)$  takes the ``universal'' form:
\begin{equation}
	\label{eq/main-results/filter}
	f_{\text{FP}}(\tau)=\frac{\varphi_{\tau_e}(\tau-\tau_L)
	\,\varphi^*_{\tau_e}(\tau)}{\langle
	\varphi_{\tau_e}|{}^{\tau_L}\varphi_{\tau_e}\rangle}
\end{equation}
where the overlap
\begin{equation}
	\label{eq/main-results/free-electrons-contrast}
\langle\varphi_{\tau_e}|
	{}^{\tau_L}\varphi_{\tau_e}\rangle
	=
\int_{\mathbb{R}}\varphi^*_{\tau_e}(t)\varphi_{\tau_e}(t-\tau_L)\,\md
t\,
\end{equation}
is the absolute contrast in the absence of interactions within the Fabry-Perot interferometer. To have an observable interference signal, the experiment is performed in a regime where $\tau_e\simeq 35\ \text{to}\ \SI{150}{\pico\second}$ which is larger than $\tau_L\simeq \SI{30}{\pico\second}$: the electronic wavepackets are wider than the Fabry-Perot loop.

The explicit expression of the filter $f_{\text{FP}}(\tau)$ is given by Eq.~\eqref{eq/uncoupled-12/FP-filter} and and its behavior in modulus and phase is discussed in details in Sec.~\ref{sec/filter-analysis}. In the experiment $\tau_L/2\tau_e\lesssim 0.5$, we have shown that $f_{\text{FP}}(\tau)$ probes the electrical phase $\me^{\phi_1(V_g,t)}$ over a time window of width $\sim \tau_e$ around $t_0$ with a smooth phase modulation. This ultimately justifies why the Fabry-Perot interferometer can be seen as a time resolved probe of this phase.

\section{Computing the interference contrast}
\label{sec/modeling}

	\subsection{Phase of the interferometer $\phi_{AB}$}
\label{sec/FP/total-phase}

Before discussing the propagation of the single electronic excitation injected by the source $S$ into the interferometer, let us comment on the phase $\phi_{\text{AB}}$ appearing in Eqs.~\eqref{interf} and \eqref{eq/interf-dc}. 

This phase corresponds to the static electromagnetic phase associated with the propagation of an electronic destruction operator along a certain closed path $\gamma$ (see Fig.~\ref{fig/FP}C). In presence of a static electrical potential $V(\mathbf{r})$ and vector potential $\mathbf{A}(\mathbf{r})$, the phase accumulated by a charge $-e$ propagating between times $t_i$ and $t_f$ is
\begin{subequations}
	\label{eq/static-phase}
\begin{align}
	\phi[\gamma]&=\frac{q}{\hbar}\int_{t_i}^{t_f}\left[\dot{\mathbf{r}}(t)\cdot
	\mathbf{A}(\mathbf{r}(t))-V(\mathbf{r}(t))\right]\,\md t\\
	&=-2\pi\,\frac{e\Phi_B}{h}+\frac{e}{\hbar}\int_{t_i}^{t_f}V(\mathbf{r}(t))\,\md
	t
\label{eq/static-phase/2}
\end{align}
\end{subequations}
in which the first term corresponds to the Aharonov-Bohm magnetic phase associated with the magnetic flux $\Phi_B$ enclosed by the curved path $\gamma$ and the second term is the electrostatic phase associated with the static component of the electrical potential experienced by the particle during its motion. In a Fabry-Perot quantum Hall interferometer, this phase depends on the precise geometry of the edge channel along which the fermionic operator $\psi_D(t)$ is propagated, as explained in \cite{Halperin-2011-1}. 

When Coulomb interactions are not taken into account, it contains a dependence on the dc voltage applied to the plunger gate $V_G^{\text{dc}}$ because the shape of the edge channel, and thus the total area enclosed by $\gamma$ depends on $V_G^{\text{dc}}$. 

When Coulomb interactions are taken into account, the second term in  the r.h.s of Eq.~\eqref{eq/static-phase/2} must be taken into account and reflect static charges present in the system. In the Fabry-Perot configuration, these are the charges stored in the closed channel appearing on Fig.~\ref{fig/FP}A. These charges are not expected to vary during an experimental run since these channels are closed. Of course, it depends on the capacitive coupling between these channels and the channels $1$ and $2$ since it determines the phase accumulated by one electron along the path $\gamma_+$ shown on Fig.~\ref{fig/FP}C. But the discussion of Ref.~\cite{Halperin-2011-1} could be applied to analyze this phase more quantitatively. The mixed dependence in the external magnetic field $B$ and the dc plunger gate voltage determines the stripped pattern of the interference contrast as a function of these two parameters. As explained in the main body of this work, the interferometer is operated in a regime where Coulomb interactions are quite strong, but not in the extreme Coulomb dominated regime where the $B$ dependence is expected to disappear \cite{Halperin-2011-1}.

This static phase being isolated, the electronic operator $\psi_D(t)$ will be back-propagated along the paths depicted on Fig.~\ref{fig/FP}C or \ref{fig:FP_2}B within the various models considered in this section: time dependent single particle scattering first and then, taking into account Coulomb interactions between charged hydro-dynamical modes (EMPs) and other dynamical degrees of freedom.

\subsection{Free electron discussion}
\label{sec/FP/no-interactions}
	
We start by considering the FP interferometer operating in a regime where Coulomb interaction effects between electrons can be neglected. They will nevertheless experience the effect of the external time dependent potential imposed by the top gate. This amounts to using a time dependent single particle scattering approach. Under this assumption, we can consider that electrons propagate within one edge channel as in Fig.~\ref{fig/FP}B. 

\subsubsection{General form of the result}
\label{sec/FP/no-interaction/intro}

In the weak-backscattering regime, we evaluate the average interference current by back-propagating the fermionic field along $\gamma_\pm$ in Fig.~\ref{fig/FP}C. When the source injects a single electron excitation in wavefunction $\varphi_S$, the interference contribution to the average current at time $t$ is given by:
\begin{equation}
	\label{eq/FP/no-interactions/1}
	{I_+(t)}=-T_A T_B\sqrt{R_A R_B} \int_{\mathbb{R}^2}	\mathcal{Z}^*_2(t-t_-)	\big(\mathcal{Z}_2 \ast R_1 \ast
	\mathcal{Z}_2\big)(t,t_+)\,\varphi_S(t_+)\varphi^*_S(\tau_-)\, \md t_+ \md t_-
\end{equation}
where the convolution is defined as 
\begin{equation}
	    \label{eq/FP/no-interactions/2}
	\big(\mathcal{Z}_2 \ast R_1 \ast
	\mathcal{Z}_2\big)(t,t')=\int_{\mathbb{R}^2}
	\mathcal{Z}_2(t_A-t') R_1(t_A,t_B)	\mathcal{Z}_2(t-t_B)\,\md t_A
	\md t_B\, .
\end{equation} 
in which $\mathcal{Z}_2(\tau)$ denotes the amplitude for propagation from $B_+$ to $A_-$ along branch $2$ in a time $\tau$ and $R_{1}(t,t')$ denotes the single particle scattering amplitude from $(1,A_+)$ to $(1,B_-)$ at respective times $t'$ and $t$. Note that due to the application of the time dependent potential $V_G$ it does depend on $t-t'$ as well as of $(t+t')/2$. By contrast, $\mathcal{Z}_2$ only depends on time differences since electrons propagating along channel $2$ do not experience the time dependent potential $V_G(t)$. The convolution $\big(\mathcal{Z}_2 \ast R_1 \ast \mathcal{Z}_2\big)(t,t')$ represents the total single particle scattering amplitude along path $\gamma_+$, up to the QPC reflection and transmission amplitudes which have been taken out for convenience.

\subsubsection{Propagation along branch $1$}
\label{sec/FP/no-interaction/branch-1}
	
The scattering amplitude $R_1(t,t')$ contains information about the phase acquired by the electrons when experiencing the influence of $V_G(t)$. The geometry of the sample suggests that the electrons only feel $V_G(t)$ beneath the top gate. Since propagation is chiral, the most general expression for $R_1(t,t')$ is
\begin{equation}
	    \label{eq/FP/no-interactions/3}
	R_1(t,t')=\int_{\mathbb{R}^2}
	R_{1\setminus G,<}(t'_G-t')R_G(t_G,t'_G)R_{1\setminus G,>}(t,t_G)
	\md t_G\,\md t'_G
\end{equation}
in which $R_{1\setminus G,<}(t'_G-t')$ (resp. $R_{1\setminus G,>}(t,t_G)$) represents the amplitude to travel across the part of branch $1$ before (resp. after) the parts beneath the top gate between times $t'$ and $t'_G$ (resp. $t_G$ and $t$). The amplitude $R_G(t_G,t'_G)$ then represents the amplitude for the particle to enter the region beneath the top gate at time $t'_G$ and exit it at time $t_G$. It depends on the time dependent top gate voltage. If we assume that propagation beneath this short top gate is ballistic with time of flight $\tau_G$ and that the electrons feel the time dependent $V_G(t)$, we have
\begin{equation}
	    \label{eq/FP/no-interactions/4}
	R_G(t_G,t'_G)=\delta(t_G-t'_G-\tau_G)\,\me^{\frac{\mi
	e}{\hbar}\int_{t'_G}^{t_G}V_G(\tau)\,\md\tau}\,.
\end{equation}
This discussion shows that because of the integration over $t_G$, the electrical potential felt by the electrons may get blurred by quantum spreading of the wavepacket during its propagation before the top gate. Such an effect would certainly limit the time resolution of the interferometer. 

We expect this dispersive blurring to be present whenever the inverse of the duration $\tau_e$ of the electronic wavepacket is of the order of the energy scale at which linear dispersion for electrons propagating within the edge channels is not constant. In the integer quantum Hall regime, this is expected to occur for $\tau_e^{-1}\gtrsim \omega_c$, where $\omega_c$ is the cyclotron frequency which is usually of the order of the terahertz: $\omega_c/2\pi\sim \SI{0.45}{\tera\hertz}\times B\,(\text{in}\ \si{\tesla})$ in AlGaAs/GaAs. This may become an issue when single electron wavepackets of duration close or even below $\SI{1}{\pico\second}$ are used, but not in the present experiment where $\tau_e\geq \SI{30}{\pico\second}$.

\subsubsection{Ballistic propagation}
\label{sec/FP/no-interaction/ballistic}

These considerations suggest that we should here restrict ourselves to ballistic propagation within the FP interferometer. In the case of branch $1$, denoting by $\tau_1$ the total time of flight and by $\tau_G$ the time of flight beneath the top gate, we have:
\begin{equation}
	    \label{eq/FP/no-interactions/5}
		R_{1\setminus G,<}(\tau)= R_{1\setminus G,>}(\tau)=\delta
		\left(\tau-\frac{\tau_1-\tau_G}{2}\right)
	\end{equation}
and using Eqs.~\eqref{eq/FP/no-interactions/3} and
\eqref{eq/FP/no-interactions/4}, this leads to
\begin{equation}
	    \label{eq/FP/no-interactions/6}
	R_{1}(t,t')=\me^{\mi\phi_1[V_G,t]}\delta(t-t'-\tau_1)\,
\end{equation}
where the phase $\phi_1[V_G,t]$ is
\begin{equation}
	    \label{eq/FP/no-interactions/7}
\phi_1[V_G,t]=\frac{e}{\hbar}
\int_{-\tau_G/2}^{\tau_G/2}
V_G\left(t-\frac{\tau_1}{2}+\tau\right)\,\md\tau\sim
	\frac{e\tau_g}{\hbar}\,V_G\left(t-\frac{\tau_1}{2}\right)
	\underset{\tau_G=R_KC_G}{=}
	2\pi\frac{C_GV_G(t-\tau_1/2)}{e}\,.
\end{equation}
The last approximation is valid in the limit of $\tau_G\ll \tau_1$ and $V_g(t)$ varying slowly over time scales $\sim\tau_G$. Ballistic propagation along branch $2$ is described by $\mathcal{Z}_2(\tau)=\delta(\tau-\tau_2)$. When a single electron excitation of wavefunction $\varphi_S$ is injected by $S$, the interference contribution to the dc average  current is of the form $I^{(\mathrm{dc})}_+=T_A T_B\sqrt{R_A
R_B} \, {X^{(\mathrm{dc})}_+}$ with
\begin{equation}
   \label{eq/FP/no-interactions/8}
[X^{(\mathrm{dc})}_+]_{V_G}=\int_{\mathbb{R}} 
	\varphi_S(t-\tau_L)\,\varphi_S^*(t)
	\,\me^{\mi\phi_1(V_G,t-\tau_2)}
\md t \,.
\end{equation}
in which $\tau_{L}=\tau_1+\tau_2$ is the total time of flight along the loop. Let us now specialize this for a Lorentzian excitation of duration $\tau_e$ is injected by $S$ at a time $t_0$:
\begin{equation}
	    \label{eq/FP/no-interactions/10}
		\varphi_S(t)=\varphi_{\tau_e}(t-t_0)\quad \text{with}\ 
	\varphi_{\tau_e}(t)=\sqrt{\frac{\tau_e}{\pi}}\,\frac{1}{t-\mi\tau_e}\,.
\end{equation}
This leads to the following expression ($\tau_{12}=\tau_1-\tau_2$ and $\tau_L=\tau_1+\tau_2$):
\begin{equation}
\label{eq/FP/no-interactions/12}
[X_+^{(\text{dc})}]_{V_G}=\int_{\mathbb{R}}\frac{\tau_e}{\pi}\,
\frac{\me^{\mi\phi_1(V_G,t_0-\tau)}}{\left(\tau+\frac{\tau_{12}}{2}\right)^2
+\left(\tau_e-\frac{\mi\tau_L}{2}\right)^2}\,\md\tau\,.
\end{equation}
In particular, for $V_g=0$, we find the vacuum baseline
\begin{equation}
\label{eq/FP/no-interactions/13}
[X_+^{(\text{dc})}]_0=\int_{\mathbb{R}}\varphi_{\tau_e}(t-\tau_L)\,\varphi_{\tau_e}^*(t)\,\md
t=\frac{2\tau_e}{2\tau_e-\mi\tau_L}
\end{equation}
which leads to the following final result
\begin{equation}
\label{eq/FP/no-interactions/relative-contrast}
\frac{[X_+^{(\text{dc})}]_{V_G}}{[X_+^{(\text{dc})}]_{0}}=
\frac{1}{\pi}\int_{\mathbb{R}}\,
\me^{\mi\phi_1(V_G,t_0-\tau)}\frac{\tau_e-\frac{\mi\tau_L}{2}}{\left(\tau+\frac{\tau_{12}}{2}\right)^2
 +\left(\tau_e-\frac{\mi\tau_L}{2}\right)^2}\,\md\tau\,.
\end{equation}
This expression shows that the relative contrast $[X_+^{(\text{dc})}]_{V_G}/[X_+^{(\text{dc})}]_{0}$, as a function of
$t_0$ appears as a convolution of the electrical phase $\me^{\mi\phi_1(V_G,t)}$ by a kernel that only depends on the geometry of the interferometer (times of flights $\tau_1$ and $\tau_2$) and of the duration $\tau_e$ of the Lorentzian pulses.  Remarkably, as we will see now, the result of Eq.~\eqref{eq/FP/no-interactions/relative-contrast} is robust to electronic decoherence.

\subsection{Electronic decoherence within each branch}
\label{sec/FP/uncoupled-branches}
\subsubsection{Modeling interactions via EMP scattering}
\label{sec/FP/uncoupled-branches/intro}

In this section, propagation of electrons within the two branches of the interferometers is modeled within the bosonization formalism, which assumes that the underlying free theory is based on electrons with a linear dispersion relation. The effect of Coulomb interactions  can then be conveniently treated within the edge-magnetoplasmon scattering formalism \cite{Safi-1995-2,Safi-1999-1}. Assuming that the electron fluid is in the linear screening regime, the corresponding EMP scattering theory is linear.

In this section, we assume that Coulomb interactions do not couple branches $1$ and $2$ of the FP interferometer. This amounts to ignoring the effects of the closed edge channels that appear on Fig.~\ref{fig/FP}A but, as will be discussed in Sec.~\ref{sec/FP/coupled-branches}, we expect their effects to be important only for $\omega\gtrsim 2\pi v/L$ where $L$ is the perimeter of the FP loop. The main hypothesis of this section is that the parts of channels $1$ and $2$ are independent EMP scatterers described by the following EMP scattering amplitudes:
\begin{subequations}
\label{eq/uncoupled-12/EMP-scattering}
\begin{align}
\label{eq/uncoupled-12/EMP-scattering/1}
b_{1,\text{out}}(\omega)&=t_1(\omega)b_{1,\text{in}}(\omega)+r_1(\omega)a_{1\text{in}}(\omega)
+\kappa_1(\omega)\,V_G(\omega)\\
b_{2,\text{out}}(\omega)&=t_2(\omega)b_{2,\text{out}}(\omega)+r_2(\omega)a_{2\text{in}}(\omega)
\label{eq/uncoupled-12/EMP-scattering/2}
\end{align}
\end{subequations}
in which the $a_{1/2}(\omega)$ are the environmental modes associated to the edge channels $1$ and $2$. There may be one or several of such modes some of which may or may not carry charge but we assume that energy conservation ensures unitarity of the total scattering matrix for these bosonic excitations. The transmission amplitudes are then related to the finite frequency admittance $Y_{11}(\omega)$ and $Y_{22}(\omega)$ via the usual expression \cite{Safi-1999-1,Degio-2010-1}:
\begin{equation}
\label{ea/uncoupled-12/S-to-Y}
Y_{\alpha\alpha}(\omega)=\frac{e^2}{h}(1-t_{\alpha}(\omega))\,
\end{equation}
whith the usual convention that the current is defined as the total current entering the region of the edge channel under consideration: $I_\alpha=i_{\alpha,\text{in}}-i_{\alpha,\text{out}}$. In a similar way, the coefficient $\kappa_1(\omega)$ describes the frequency dependent linear response of the edge current $i_{11,\text{out}}(\omega)$ to $V_G(\omega)$. More precisely
\begin{equation}
\label{eq/uncoupled-12/EMP-scattering/kappa1-Y}
Y_{1,G}(\omega)=e\sqrt{\omega}\,\kappa_1(\omega)
\end{equation}
Under the hypothesis of the present subsection, edge channel $2$ does not respond to $V_G(t)$ when the QPC are opened and this is why there is no linear term involving $V_G(\omega)$ in Eq.~\eqref{eq/uncoupled-12/EMP-scattering/2} contrary to Eq.~\eqref{eq/uncoupled-12/EMP-scattering/1}.
	
\subsubsection{Results}
\label{sec/FP/uncoupled-branches/results}
	
As in the previous section the idea is to back-propagate the fermionic field $\hat{\psi}_D(t)$ along the two main paths in order to connect the FP geometry to the MZ formalism in the same spirit as in Ref.~\cite{Souquet-Basiege-2024-1}. When a single electron excitation with wavefunction $\varphi_S$ is injected, the average time-dependent current is then obtained as
\begin{equation}
	\label{eq/uncoupled-12/1}
	{I_+(t)}=-T_A \sqrt{R_A R_B} \,\me^{2\mi\theta_2}
	\me^{\mi\phi_1(V_G,t)}  \int \md t_+ \md t_- \, \mathcal{Z}_{2}(t_-
	-(t-\tau_2))\,\mathcal{Z}_1(t-\tau_2-t_+)\,
	\varphi_S^*(t_-)\varphi_S(t_+)
\end{equation}
where the phase $\phi(V_G,t)$ appears as the convolution of the gate voltage by a kernel which describes the filtering associated with the capacitive coupling to the top gate. More specifically, introducing the finite frequency admittance of the dipole formed by the plunger gate and the branch $1$ of the FP interferometer:
\begin{subequations}
	\label{eq/uncoupled-12/Frank-Condon}
	\begin{align}
		\label{eq/uncoupled-12/Frank-Condon/1}
	\phi_1(V_G,t) &= \frac{e}{\hbar}\left(\Gamma_{1,G}\ast V_G\right)(t)\\
	\widetilde{\Gamma}_{1,G}(\omega) &= \frac{R_KY_{1,G}(\omega)}{-\mi\omega}
	\label{eq/uncoupled-12/Frank-Condon/2}
	\end{align}
\end{subequations}
In principle, one should therefore use a model of electrostatics of the system to derive this finite frequency admittance, for example in the spirit of the discrete element modeling of a top gate capacitively coupled to an edge channel in Ref.~\cite{Souquet-Basiege-2024-1}. But due to the geometry of the sample considered here, one can use the expression obtained in Eq.~\eqref{eq/FP/no-interactions/7}.  Note that the phase
\begin{equation}
\label{eq/uncoupled-12/2}
\theta_2=\mathrm{Im}\bigg[\int_0^{+\infty} \frac{\md\omega}{\omega} 
\big( t^*_2(\omega)\me^{\mi\omega \tau_2}-1 \big) \bigg]\,.
\end{equation}
does not depend on time.
The two amplitudes $\mathcal{Z}_1(\tau)$ and $\mathcal{Z}_2(\tau)$ correspond to elastic scattering amplitude for a single electron excitation on top of the Fermi sea propagating across branches $1$ and $2$ respectively. As in Ref.~\cite{Souquet-Basiege-2024-1}, they are expressed in terms of the elastic scattering amplitudes for energy resolved single electron excitations $\widetilde{Z}_\alpha(\omega>0)$ via a Fourier transform (see Refs.~\cite{Degio-2009-1,Cabart-2018-1}):
\begin{equation}
\label{eq/uncoupled-12/3}
\mathcal{Z}_{\alpha}(\tau)=\int_0^{+\infty}\widetilde{\mathcal{Z}}_\alpha(\omega)\,\me^{-\mi\omega\tau}\,
\frac{\md \omega}{2\pi}\,
\end{equation}
and therefore contain not only the information about the Wigner-Smith time delay for low frequency excitations but also about electronic decoherence along the branches $1$ and $2$ of the Fabry-Perot interferometer. Putting all these results together leads to the following expression
\begin{equation}
\label{eq/uncoupled-12/X+-time}
\left[X_+^{(\text{dc})}\right]_{V_G}=e^{2i\theta_2}\int_{\mathbb{R}^3}
\me^{\mi\phi_1(V_G,t-\tau_2)}\mathcal{Z}_{2}(t_--
(t-\tau_2))\,\mathcal{Z}_1(t-\tau_2-t_+)\,
\varphi_{S}^*(t_-)\varphi_{S}(t_+)
\md t_+\md t_-\md t
\end{equation}
which, for $\varphi_S(t)=\varphi_{\tau_e}(t-t_0)$ in which $\varphi_{\tau_e}$ defined by Eq.~\eqref{eq/Leviton-wavefunction}, can then be rewritten as a filtering
\begin{equation}
\label{eq/uncoupled-12/filtering}
\left[X_+^{(\text{dc})}\right]_{V_G}=\me^{2\mi\theta_2}\int_{\mathbb{R}}
\widetilde{f}_{\text{FP}}(\Omega)\widetilde{\mathcal{F}}_{V_G}(\Omega)\,\me^{-\mi\Omega t_0}
\,\frac{\md\Omega}{2\pi}
\end{equation}
of $\mathcal{F}_{V_G}(t)=\me^{\mi\phi_1(V_g,t)}$ by the filter
\begin{equation}
\label{eq/uncoupled-12/FP-filter}
\widetilde{f}_{\text{FP}}(\Omega)=4\pi \tau_e\int_{|\Omega|/2}^{+\infty}\me^{-2\omega\tau_e}
\widetilde{\mathcal{Z}}_1\left(\omega-\frac{\Omega}{2}\right)\,
\widetilde{\mathcal{Z}}_2\left(\omega+\frac{\Omega}{2}\right)
\frac{\md\omega}{2\pi}\,.
\end{equation}
Note that here, the Lorentzian shape of the current pulse is responsible for the $\me^{-\omega\tau_e}$ in the r.h.s. of Eq.~\eqref{eq/uncoupled-12/FP-filter}: it reflects the Leviton's exponentially decaying wavefunction in energy.  Finally, the vacuum baseline is given by
\begin{equation}
\label{eq/uncoupled-12/vacuum-baseline}
\left[X_+^{(\text{dc})}\right]_0=\widetilde{f}_{\text{FP}}(\Omega=0)=
4\pi \tau_e\int_{0}^{+\infty}\me^{-2\omega\tau_e}
\widetilde{\mathcal{Z}}_1(\omega)\,
\widetilde{\mathcal{Z}}_2(\omega)
   \frac{\md\omega}{2\pi}\,
\end{equation}
Note that, as expected, this expression as the same form as in Ref.~\cite{Souquet-Basiege-2024-1} with $\mathcal{Z}_1(\omega)\mathcal{Z}_2(\omega)$ playing the role of $\mathcal{Z}_1(\omega)$.

\subsubsection{Adiabatic approximation}

It is then convenient to isolate the contribution of the Wigner-Smith time delays $\tau_{1,2}$ for the EMP modes respectively propagating along branches $1$ and $2$ of the FPI,  by rewriting
\begin{equation}
\label{eq/uncoupled-12/Wigner-Smith-out}
\widetilde{\mathcal{Z}}_\alpha(\omega)=\me^{\mi\omega\tau_\alpha}
\widetilde{\mathcal{Z}}^{(0)}_\alpha(\omega)
\end{equation}
in which $\widetilde{\mathcal{Z}}^{(0)}_\alpha(\omega)$ contains all the effects of decoherence associated with the dispersion and scattering of the EMP modes ($\widetilde{\mathcal{Z}}^{(0)}_\alpha(\omega)=1$ for ballistic propagation with time of flight $\tau_\alpha$). The filter function $f_{\mathrm{FP}}(\Omega)$ can then be rewritten as
\begin{equation}
\label{eq/uncoupled-12/FP-filter/rewritten}
\widetilde{f}_{\text{FP}}(\Omega)=4\pi \tau_e \me^{-|\Omega|\tau_e}
\me^{\mi(|\Omega|\tau_L-\Omega\tau_{12})}\int_{0}^{+\infty}\me^{-2\omega\tau_e}
\me^{\mi\omega\tau_L}
\widetilde{\mathcal{Z}}_1^{(0)}\left(\omega+\frac{|\Omega|-\Omega}{2}\right)\,
\widetilde{\mathcal{Z}}_2^{(0)}\left(\omega+\frac{|\Omega|+\Omega}{2}\right)
   \frac{\md\omega}{2\pi}\,.
\end{equation}
Exactly as in Ref.~\cite{Souquet-Basiege-2024-1}, we perform an adiabatic approximation by assuming that, whenever the frequencies involved in $\mathcal{F}_{V_G}(t)$ are much lower than the frequencies at which the amplitudes $\widetilde{\mathcal{Z}}^{(0)}_\alpha$ vary, then the $\Omega$ dependance in their argument in the r.h.s. of Eq.~\eqref{eq/uncoupled-12/FP-filter/rewritten} can be neglected. This leads to
\begin{equation}
\label{eq/uncoupled-12/FP-filter/adiabatic}
\widetilde{f}_{\text{FP}}(\Omega)=e^{-|\Omega|\tau_e}
e^{i(|\Omega|\tau_L-\Omega\tau_{12})/2}\left[X_+^{(\text{dc})}\right]_0
\,.
\end{equation}
Substituting this expression in the filtering equation \eqref{eq/uncoupled-12/filtering} enables us to obtain the relative contrast as a convolution:
\begin{equation}
\frac{[X_+^{(\text{dc})}]_{V_G}}{[X_+^{(\text{dc})}]_0}=
\int_{\mathbb{R}}f_{\text{FP}}(\tau)\,\mathcal{F}_{V_G}(t_0-\tau)\,\md\tau
\end{equation}
in which the convolution kernel is obtained as
\begin{equation}
f_{\text{FP}}(\tau)=\frac{1}{\pi}\,
\frac{\tau_e-\frac{\mi\tau_L}{2}}{\left(t+
\frac{\tau_{12}}{2}\right)^2+\left(\tau_e-\frac{\mi\tau_L}{2}\right)^2}
\end{equation}
This is  exactly the same expression as in Eq.~\eqref{eq/FP/no-interactions/relative-contrast}. This proves that, in the absence of coupling between the two branches and when the adiabatic approximation is valid, the filtering of the time dependent voltage is identical to the one obtained via time dependent single particle scattering (see Sec.~\ref{sec/FP/no-interactions}), up to renormalization of the vacuum baseline due to electronic decoherence.

Such a result could indeed be expected since the adiabatic approximation means that information about the time dependent phase $\mathcal{F}_{V_G}(t)$ is stored into edge-magnetoplasmon modes that are not too much affected by dispersion and dissipation: their transmission amplitudes across branches $1$ and $2$ are close to the ballistic amplitudes $\me^{\mi\omega\tau_{1,2}}$. This explains why electronic decoherence is unaffected by the presence of the time dependent voltage $V_G(t)$.  Note however that the theory presented here enables, in principle, to account for non-adiabatic effects by using Eq.~\eqref{eq/uncoupled-12/FP-filter/rewritten} instead of Eq.~\eqref{eq/uncoupled-12/FP-filter/adiabatic}.
	
\subsection{Coupling between the two branches}
\label{sec/FP/coupled-branches}

In the previous section, Coulomb interactions have been introduced under the hypothesis that the upper and lower branches $1$ and $2$ of the FP interferometer are not coupled electrostatically. However, due to the presence of closed edge channels on Fig.~\ref{fig/FP}A, this is not true. Here, we discuss the  effects of Coulomb interactions in the presence of such closed loops. Let us stress that the effect of the total (static) charge stored on the closed channels has already been incorporated in the effective Aharonov-Bohm phase and therefore, we are only discussing the effect of Coulomb interactions on the so-called hydrodynamic (EMP) modes.

This leads to a more complex scattering matrix for the edge-magnetoplasmon modes of the upper branch $\hat{b}_1$ and of the lower one $\hat{b}_2$. We have to introduce scattering amplitudes $t_{12}(\omega)$ and $t_{21}(\omega)$ respectively connecting $\hat{b}_{1,\text{out}}(\omega)$ to $\hat{b}_{2,\text{in}}(\omega)$ and $\hat{b}_{2,\text{out}}(\omega)$ to $\hat{b}_{1,\text{in}}(\omega)$. These EMP transmission coefficients are related to the finite frequency admittances of the edge channels when the QPCs are fully opened via
\begin{equation}
Y_{\alpha\beta}(\omega)=\frac{e^2}{h}\left(1-t_{\alpha\beta}(\omega)\right)\,.
\end{equation}
Then, the $\hat{b}_{2,\text{out}}(\omega)$ modes now responds to the gate voltage $V_G(\omega)$ through a coefficient $\kappa_2(\omega)$ associated with the finite frequency admittance $Y_{2,G}(\omega)$. Finally, exactly as before, there may be other environmental modes to ensure charge conservation as well as energy conservation so that the full scattering matrix for all these bosonic degrees of freedom is unitary.

A crude estimate can be used at low frequencies to obtain an order of magnitude of the finite frequency admittance between the two branches $1$ and $2$ of the FP interferometer. First of all, folding each of these branches together enables us to see them as a transmission line of characteristic impedance $R_K/2$. They are capacitively coupled to a closed loop which forms a quantum Hall Fabry Perot interferometer (see Ref.~\cite{Frigerio-2024-1} for a similar but slightly different geometry) for EMP modes propagating associated with the $\nu_{\text{FP}}-1=2$ inner channels. At low frequency, such an interferometer can be roughtly seen as the series addition of two capacitances and an RL circuit where $R= R_K/2(\nu_{\text{FP}}-1)$ and $L=2R\,\tau_{RL}$ where $\tau_{RL}$ is the characteristic $RL$ time associated with a quantum Hall bar \cite{Delgard-2021-1} at filling fraction $\nu_{\text{FP}}-1$ whose length is of the order of the cavity's perimeter (up to geometric factors). In the end, the finite frequency admittance of such a dipole is of the order
\begin{equation}
	R_KY(\omega)=-\mi(\nu_{\text{FP}}-1)\,\frac{\omega}{\omega_{\text{LC}}}+\mathcal{O}
	\left(\frac{\omega}{\omega_{\text{LC}}}\right)^2\,
\end{equation}
in which $\omega_{\text{LC}}=1/\sqrt{LC}$ is the resonance frequency of the $LC$ resonator which, here, corresponds to the lowest resonance frequency of the EMP cavity formed by the $\nu_{\text{FP}}-1$ closed edge channels of the FP interferometer. This explains why, at low frequencies compared to this first resonance frequency, the coupling between the edge channel $1$ and $2$ is indeed small.

Of course, it would be interesting to account for the effect of this capacitive coupling mediated by the closed edge channels (see  Ref.~\cite{Wei-2024-1} in the stationary regime) but we think that the experimental data clearly show that it is not relevant for the experiment presented in this paper. 

\section{Filtering of $\mathcal{F}_{V_G}(t)$}
\label{sec/filter-analysis}

Let us now discuss the time resolution associated with Leviton wavepackets of duration $\tau_e$ which follows from the filter $f_{\text{FP}}(\tau)$ whose expression is, for $\tau_1=\tau_2$ (as in the sample under consideration):
\begin{equation}
f_{\text{FP}}(\tau)=\frac{1}{\pi}\,\frac{\tau_e-\frac{\mi\tau_L}{2}}{\tau^2
 +\left(\tau_e-\frac{\mi\tau_L}{2}\right)^2}\,.
\end{equation} 
Analyzing the modulus of this function as a function of $\tau$ shows that there are two distinct regimes: the first one which corresponds to \emph{long wavepackets} ($\tau_e<\tau_L/2$) and the regime of \emph{short wavepackets} ($\tau_e>\tau_L/2$).

The regime of short wavepackets is the regime of interest in the present experiment since, in order to observe a significant contrast even for $V_G(t)=0$, one has to chose $\tau_L\lesssim \tau_e$ since, for example, when interactions can be neglected, the vacuum baseline whose modulus gives the contrast of interference fringes reduces to:
\begin{equation}
\left[X_+^{(\text{dc})}\right]_0=\frac{\tau_e}{\tau_e-\frac{\mi\tau_L}{2}}\,.
\end{equation}

\subsection{Long wavepackets}
\label{sec/filter-analysis/long-wp}

\begin{figure}
\centering
\includegraphics[width=14cm]{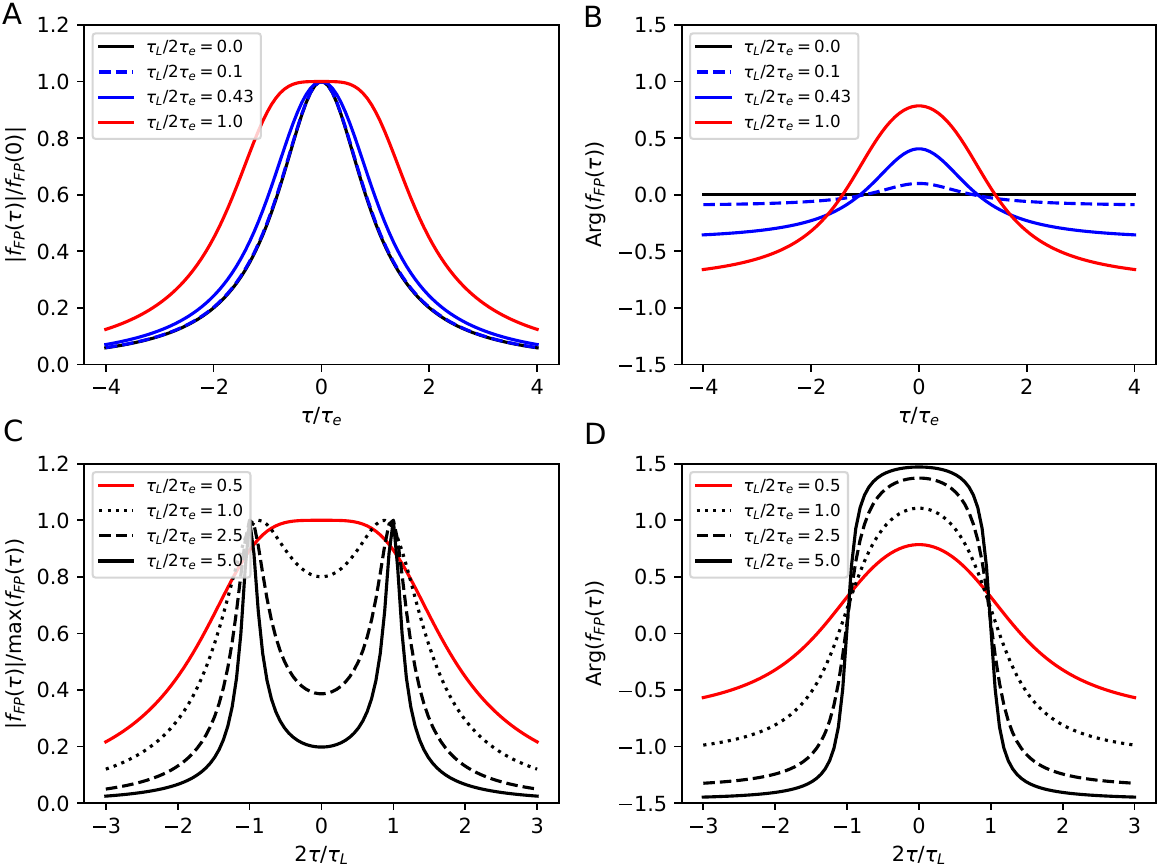}

\caption{\label{fig/filter-analysis} {\bf (A) and (B)}: plot of $|f_{\text{FP}}(\tau)/f_{\text{FP}}(0)|$ and $\mathrm{Arg}(f_{\text{FP}}(\tau))$ as a function of $\tau/\tau_e$ for long wave packets: $\tau_L/2\tau_e\rightarrow 0$ (grey line), $\tau_L/2\tau_e=1/10$ (blue dashed line), $0.43$ (blue line), $1$ (red line). The blue curves correspond to the experimentally realized $\tau_e$ ranging from $\SI{35}{\pico\second}$ to $\SI{150}{\pico\second}$ with $\tau_L=\SI{30}{\pico\second}$. {\bf (C) and (D)}: Plot of $|f_{\text{FP}}(\tau)|/\max_\tau(|f_{\text{FP}}(\tau)|)$ and of $\mathrm{Arg}(f_{\text{FP}}(\tau))$ as a function of $2\tau/\tau_L$ for short wavepackets: $\tau_L/2\tau_e=1$ (red line), $2$ (black dotted line), $5$ (black dashed line) and $10$ (black line).}
\end{figure}

In this regime, $|f_{\text{FP}}(\tau)|$ has its maximum for $\tau=0$ and then decays to zero for $|\tau|/\tau_e\gg 1$. As shown on Fig.~\ref{fig/filter-analysis}A, as soon as $\tau_e\gtrsim \tau_L$, the filter function has a width of the order of $\tau_e$. The study of its phase, depicted on Fig.~\ref{fig/filter-analysis}B, shows that when $\tau_e\gg \tau_L/2$, it is almost constant over the interval over which the modulus takes significant values (typically $|\tau/\tau_e|\lesssim 2$). This shows that, in this limit, the filter function $f_{\text{FP}}(\tau)$ tends to ``average'' the phase factor $\mathcal{F}_{V_G}(t)$ over a window of width $\tau_e$ around $t_0$ in Eq.~\eqref{eq/main-results/1}. 

For the values considered in the present work, this is not exactly the case, especially when considering the shortest wavepackets. Nevertheless, the model considered here enables us to account quantitatively of the deviation from the simple image of a simple averaging of the phase.

\subsection{Short wavepackets}
\label{sec/filter-analysis/short-wp}

When considering short wavepackets ($\tau_e<\tau_L/2$), the behavior of $f_{\text{FP}}(\tau)$ changes drastically. First of all the modulus $|f_{\text{FP}}(\tau)|$ has two maxima for $\tau\simeq \pm\tau_L/2$ which are associated with peaks of width $\sim \tau_e$ as can be seen on  Fig.~\ref{fig/filter-analysis}C. The phase of $f_{\text{FP}}(\tau)$ is also different: as depicted on Fig.~\ref{fig/filter-analysis}D, it starts from $-\arctan(\tau_L/2\tau_e)$ before the two peaks ($-\infty <\tau <-\tau_L/2$) and then switches to $\arctan(\tau_L/2\tau_e)$ between the peaks and goes back to the previous negative value when $\tau$ increases above $\tau_L/2$. The transition happens over a duration $\sim \tau_e$. 

The appearance of the phase kink for short electronic wavepackets on Fig.~\ref{fig/filter-analysis}D is reminiscent of Ref.~\cite{Gaury-2014-1} where the effect of the phase jump induced by a Lorentzian pulse of duration much shorter than the time of flight across a coherent electronic interferometer (FP or MZI) is discussed. Nevertheless, Ref.~\cite{Gaury-2014-1} discusses the influence of the many-body phase jump which is $2\pi \overline{n}$ for a current pulse of charge $-e\bar{n}$ (see Eq.~\eqref{eq/main-results/2} here).  In particular, it discusses the behavior as a function of $\bar{n}$. By contrast, in the present work, $\bar{n}=1$ and, since we are in the IQH regime, we are discussing the effect of the phase imprinted on the single electronic excitation on top of the Fermi sea. Its wavefunction also exhibits a phase jump of the same shape but with amplitude $\pi$ (not $2\pi$, see Eq.~\eqref{eq/Leviton-wavefunction}).  Considering the a many-body phase is indeed important for probing many-body effects and in particular, discussing  the role of anyonic statistics of fractionally charged current pulses in the IQH regime \cite{June-Young-2020-1,Morel-2022-1}. But in  the present work, this is not what we are interested in: because we are using one single electron excitation as a probe of the time dependent voltage, we are considering the effects of the time dependent phase and modulus of the Leviton electronic excitation as a function of the ratio its duration $\tau_e$ to $\tau_L$.

\section{Measurement setup and processes}
    \subsection{Fridge setup}

\begin{figure}
    \centering
    \includegraphics[width = 0.4\textwidth]{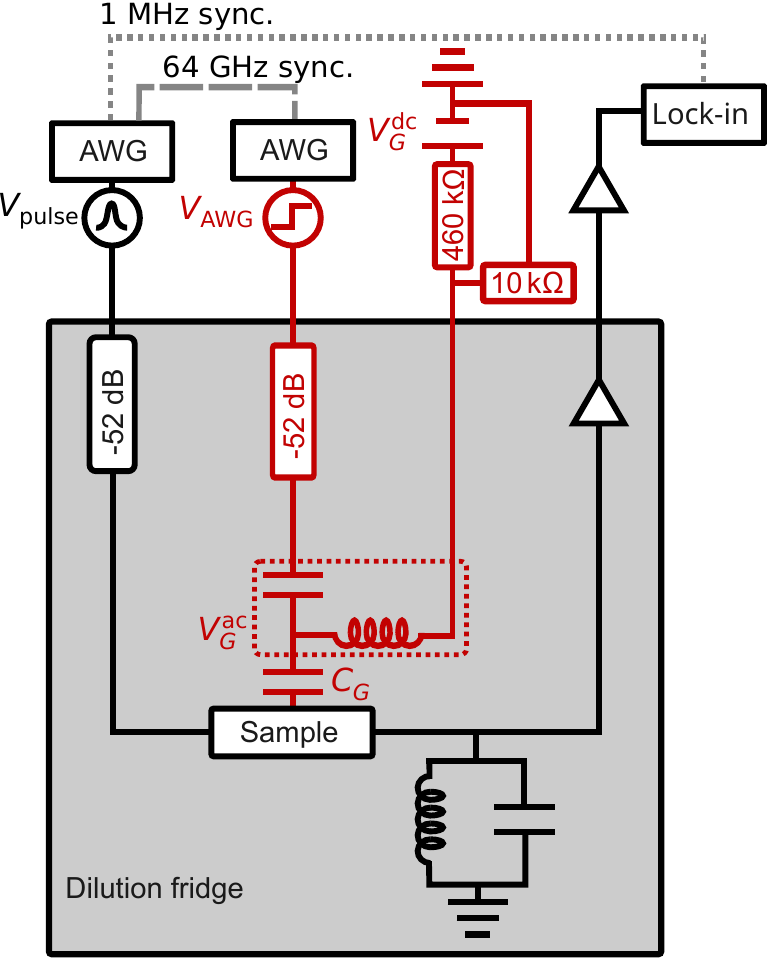}
    \caption{\textbf{Fridge setup:} The sample is placed at the bottom of a dilution refrigerator. It is connected to two AWG outputs through attenuators distributed over the various stages of the fridge for an total attenuation of \SI{-52}{\decibel}. The line connected to the gate is also connected in dc via a bias-tee where a voltage divider is present. The output signal is filtered by a tank circuit and then amplified at two stages and acquired by a lock-in amplifier.}
    \label{figS6}
\end{figure}

All the measurements presented in this paper were performed at the bottom of a dilution cryostat at base electronic temperature \SI{30}{\milli\kelvin}. The rf lines used to send the pulse and square drive to the sample are represented in figure \ref{figS6}. Both rf lines are attenuated throughout the descent to low temperature by a total of \SI{-52}{\decibel}. The gate line is also connected in dc via a bias tee placed at the level of the mixing chamber. Additionally, a voltage divider is placed inside the fridge on the dc part of the gate. The gate line is capacitively coupled to the sample through the capacitance $C_G$. 

The output signal for the measurement of noise $\Delta S$ or output current $I_{\mathrm{out}}$ passes through a tank circuit and is collected by a cryogenic amplifier before being amplified at room temperature. The tank circuit is a band-pass filter for the measurement of the noise at a frequency of $1.1$ MHz in a $100$ kHz bandwidth. It prevents the noise measurement to be polluted by unavoidable low frequency parasitic contributions.

The output current $I_{\mathrm{out}}$ is measured by a lock-in amplifier by applying a square modulation to the voltage excitation $V_{\mathrm{pulse}}(t)$ used to generate the periodic train of single electron pulses. The modulation is performed at \SI{1}{\mega\hertz}, thus averaging over many pulses generated with a \SI{1}{\giga\hertz} frequency, alternating sign at  \SI{1}{\mega\hertz}.

\subsection{Pulse calibration}
\label{appendix/pulse_calibration}

\begin{figure}[h!]
    \centering
    \includegraphics[width = \textwidth]{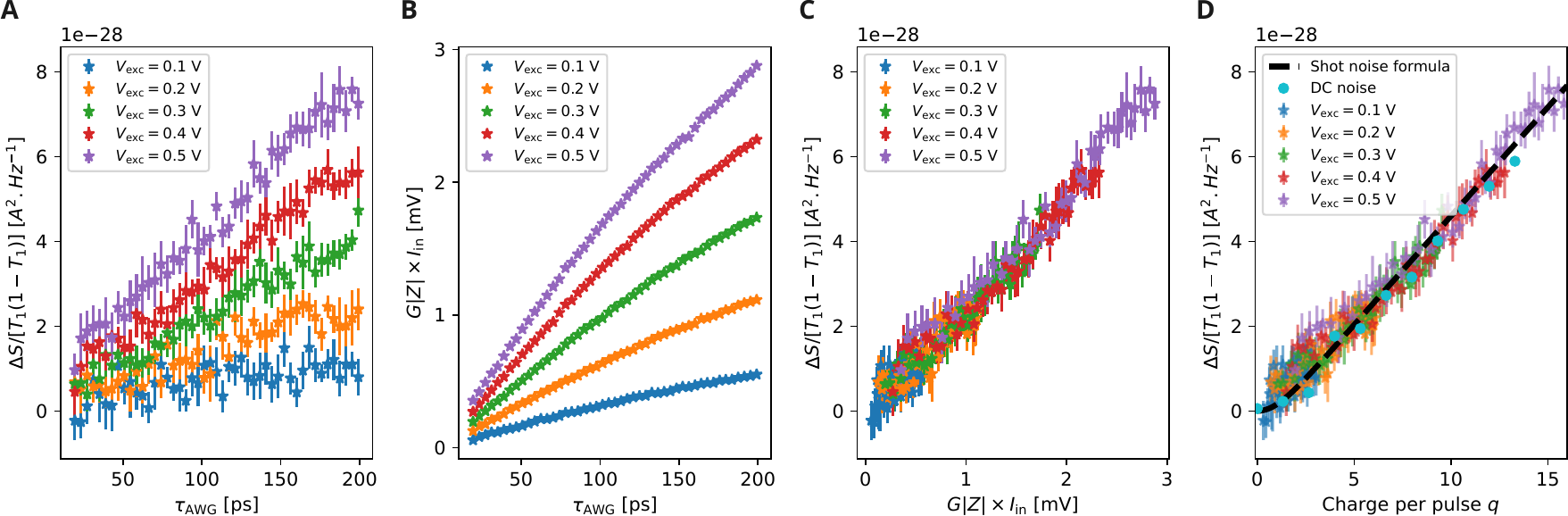}
    \caption{\textbf{Pulse calibration:} (\textbf{A}) Shot noise measured in the configuration presented in figure 1.D of the main text as a function of the width $\tau_{\mathrm{AWG}}$ of the pulses for various amplitudes $V_\mathrm{exc}$ of the voltage pulse generated by our AWG.  (\textbf{B}) Measurement of $|Z| I_{\mathrm{in}}$ as a function of the  width $\tau_{\mathrm{AWG}}$ and the excitation amplitude $V_{\mathrm{exc}}$
    (\textbf{C}) Same data as in A and B, plotting the noise $\Delta S$ as a function of $|Z| I_{\mathrm{in}}$. (\textbf{D}) Same data as in C, rescaling $|Z| I_{\mathrm{in}}$ by the factor $\alpha=1.81\cdot 10^{-4}$ V,  such that the horizontal axis now represents the total charge contained within a single pulse, $q=|Z| I_{\mathrm{in}}/\alpha$. The dashed line represents the expected behavior for shot noise and the blue dots the DC noise.}
    \label{figS1}
\end{figure}

In our experiment, single electron pulses are generated by applying a periodic train of Lorentzian pulses $V_\mathrm{pulse}(t)$ containing 1 electron. More generally, a train of Lorentzian pulses containing the charge $q$ (in units of the electron charge) and of temporal width $\tau_e$ can be written as $V_{\mathrm{pulse}}(t)= \sum_n \frac{q h \tau_e}{\pi e}\frac{1}{(t-n/f)^2+\tau_e^2}$.  $V_{\mathrm{pulse}}(t)$ is generated at room temperature by an arbitrary wave generator (AWG) generating the time-dependent voltage :
\begin{eqnarray}
V_{\mathrm{pulse,AWG}}(t) & = &  \sum_n V_{\mathrm{exc}} \frac{\tau_{\mathrm{samp}}^2}{(t-n/f)^2+\tau_{AWG}^2}, 
\label{VAWG}
\end{eqnarray} 
where $\tau_{\mathrm{samp}}=$~\SI{15.6}{\pico\second} is the sampling time of the AWG. $V_{\mathrm{pulse,AWG}}(t)$ is then attenuated at each stage of the fridge, requiring a proper calibration between the applied voltage at room temperature characterized by $V_{\mathrm{exc}}$ and $\tau_\mathrm{AWG}$ and the charge per pulse $q$. The calibration of $q$ is performed together with the calibration of the input dc current $I_\mathrm{in}$, where $I_\mathrm{in}$ is the dc component of $e^2/hV_{\mathrm{pulse}}(t) $. $I_\mathrm{in}$ is converted to a voltage on the output impedance $Z$ of the sample that consists in the Hall resistance in parallel with the LC tank circuit used for the noise measurements. It is then amplified by cryogenic and room temperature amplifiers (see Fig.\ref{figS6}) with total gain $G$. It is measured by a lock-in amplifier by applying a square modulation of $V_{\mathrm{pulse,AWG}}(t)$ at the low frequency of 1 MHz. The whole measurement process of $I_\mathrm{in}$ (modulation, conversion to a voltage on the frequency dependent impedance $Z(\omega)$, gain of the amplifiers) requires a proper calibration. 

This is performed by measuring both the calibrated excess current noise $\Delta S$ generated by the partitioning of the train of pulses at QPC1 and the uncalibrated dc current $I_{\mathrm{in}}$ for different values of the parameters $V_{\mathrm{exc}}$ and $\tau_\mathrm{AWG}$. Our measurements of the shot noise $\Delta S$ normalized by the binomial factor $T_1(1-T_1)$ (where $T_1$ is the transmission of QPC A) and of the amplified current $G |Z| \times I_\mathrm{in}$ are plotted on Figs.\ref{figS1}A and B. Both the noise and the current show a linear increase for $\tau_\mathrm{AWG} \leq 100$ ps. This is expected as in the limit of non-overlapping pulses, $q$ is expected to depend linearly on both 
 $\tau_\mathrm{AWG}$ or $V_{\mathrm{exc}}$. For larger values of  $\tau_\mathrm{AWG}$, a sublinear variation of $\frac{\Delta S}{T_1(1-T_1)}$ and $G |Z| I_\mathrm{in}$ is observed that reflects the overlap of consecutive pulses in the train.  
 
 We plot on  Figs.\ref{figS1}C the noise $\frac{\Delta S}{T_1(1-T_1)}$ as a function of  $G|Z| I_\mathrm{in}$. Remarkably, all points fall on a linear slope. This reflects that the shot noise is  proportional to the input current : $\frac{\Delta S}{T_1(1-T_1)} = 2 e I_\mathrm{in} = 2e^2 f q$, where $f=1$ GHz is the repetition frequency. We can thus calibrate the lever arm $\alpha$ relating our measurement of the input current to the charge $q$, $\alpha= G|Z|I_\mathrm{in}/q$. By choosing $\alpha=1.81\cdot 10^{-4}$ V, we impose that our data $\Delta S$ have the expected slope $2e^2 f$ when plotted as a function of $q= G|Z|I_\mathrm{in}/\alpha$ (see Fig.\ref{figS1}D). This provides both a calibration of the charge per pulse $q$ and of the input current $I_{\mathrm{in}}$.
 
 As a check of the soundness of our calibration procedure, we also plot on Fig.\ref{figS1}D the noise  $\Delta S$ generated by a dc voltage bias $V_{\mathrm{dc}}$, with $q_{\mathrm{dc}}=e V_{\mathrm{dc}}/(hf)$.  As can be seen on the figure, all our measurements fall nicely on the expected slope for shot noise $\Delta S = 2 e^2f q$. The dashed line represents the shot noise formula taking into account the temperature: $\Delta S= 2 e^2f q \times [ \coth(\frac{hfq}{2k_B T_\mathrm{el}} - \frac{2k_B T_\mathrm{el}}{hfq})] $ and using the temperature $T_\mathrm{el}=25$ mK.

\subsection{Full experimental data set}
\label{appendix/full_data_set}

\begin{figure*}[h!]
    \centering
    \includegraphics[width = \textwidth]{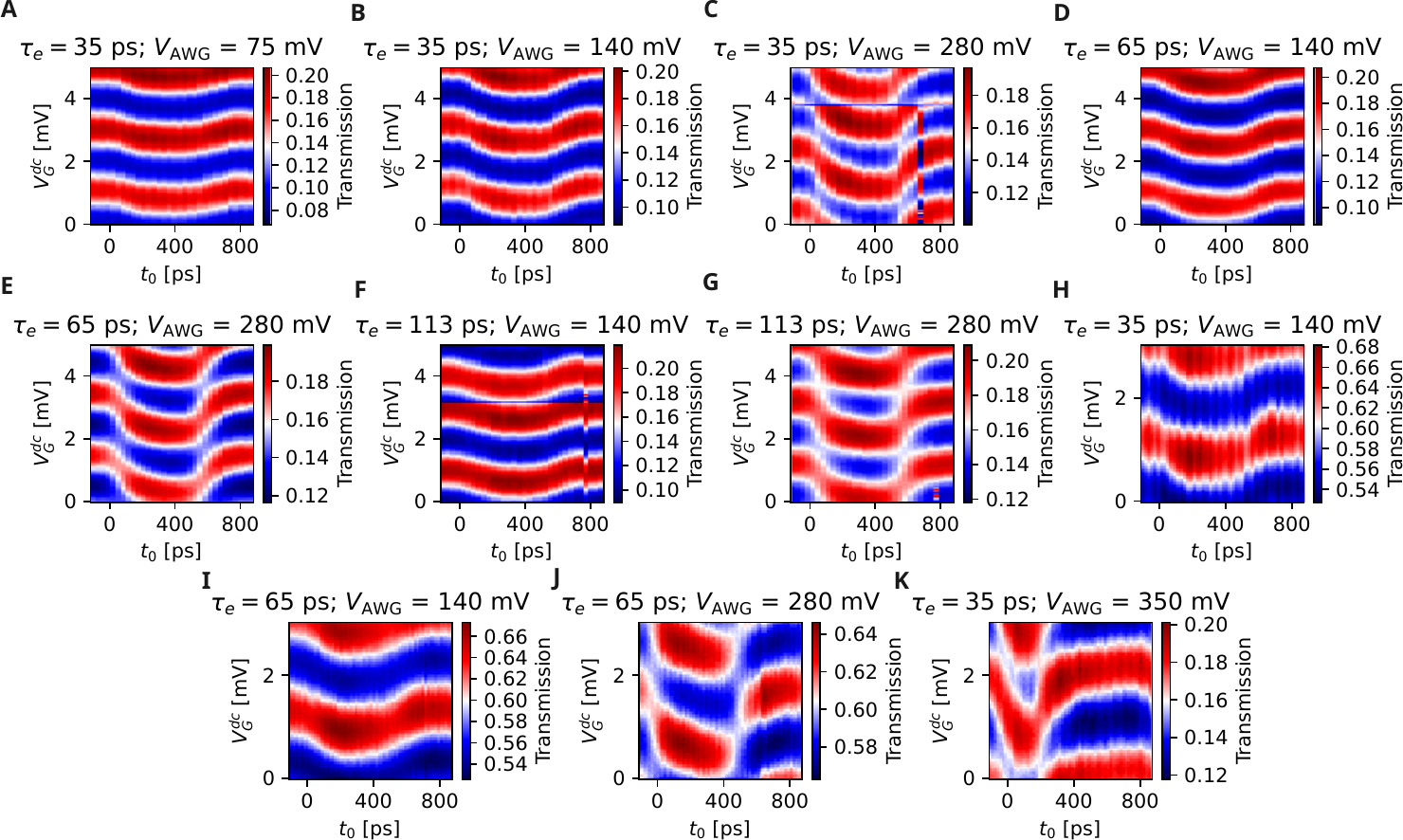}
    \caption{\textbf{Full experimental data set:} (\textbf{A})-(\textbf{K}) Time vs gate transmission maps. The pulse widths $\tau_e$ and square drive height $V_\mathrm{AWG}$ are indicated above each map. The data shown in the main text is extracted from maps A, B, C, D, E and G. Maps (\textbf{H})-(\textbf{K}) were acquired with a shorter time resolution on $t_0$. Map (\textbf{K}) was obtained by substituting the square drive by a rectangular one of width \SI{250}{\pico\second}.}
    \label{figS2}
\end{figure*}

In the main text of the article we only present three maps of the measurement of  $T(V_G^\mathrm{dc},t_0)$ . However more data was used to extract the points in figures 3I and J from the main text. The full data set used for these figures as well as additional data points are shown in figure \ref{figS2}.

\begin{figure}[h!]
    \centering
    \includegraphics[width = 0.5\textwidth]{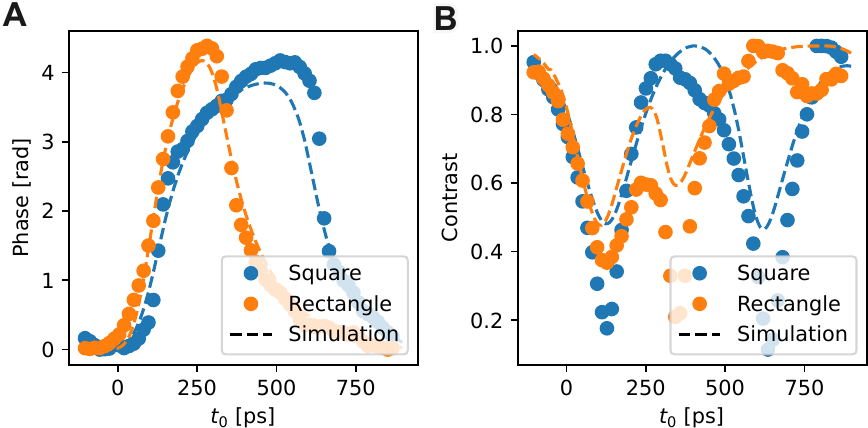}
    \caption{\textbf{Rectangle versus square drive:} (\textbf{A}) Phase extracted from maps J (square) and K (rectangle) of figure \ref{figS2}. (\textbf{B}) Associated contrast variation. In both graphs, the simulated data is shown in dashed lines.}
    \label{figS8}
\end{figure}

In particular, figure \ref{figS2}K was obtained by applying a rectangular drive (with a temporal width $\tau_s=250$ ps) instead of a square one on $V_G^\mathrm{ac}(t)$. As a result, we observe that the transmission changes on a shorter time scale. Figure \ref{figS8} presents the extracted phase and amplitude of the transmission extracted from figure \ref{figS2}J and K. On this figure the dots represent the experimental data while the dashed lines represent the simulation performed using the same parameters as in the main text, but adapting it to the physical parameters used to measure figures \ref{figS2}J and K. It should be noted here that the amplitude of the rectangle is larger (\SI{350}{\milli\volt}) that for the square (\SI{280}{\milli\volt}). However, we observe that the phase shift (as seen in figure \ref{figS8}A) does not change in the same proportions, probably due to the rise time that becomes comparable with the temporal width of the rectangular drive ($\tau_s=$~\SI{250}{\pico\second}). The measured variations of the phase are well reproduced by the model both for the square (blue dashed line, $\tau_s = \frac{1}{2f}$) and rectangle  (orange dashed line, $\tau_s = \frac{1}{4f}$) drives. While the quantitative agreement is not as good for the contrast as for the phase, we still reproduce the contrast dips in the correct position.

\subsection{Square drive calibration}

\begin{figure}[h!]
    \centering
    \includegraphics[width = 0.5\textwidth]{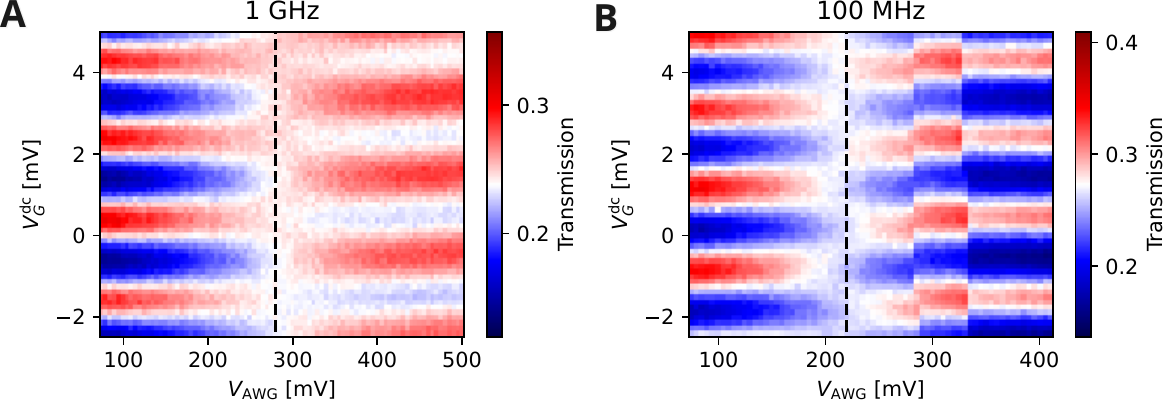}
    \caption{\textbf{Power dependence of the dc transmission:} (\textbf{A}) Dependence of the transmission of the FPI as a function of the amplitude of the square drive $V_\mathrm{AWG}$ and the dc voltage applied on the central gate for a periodic signal at \SI{1}{\giga\hertz} frequency. The vertical dashed line indicates the point where the oscillations disappear at $V_\mathrm{AWG} =$~\SI{280}{\milli\volt}, meaning that the amplitude of the square drive compensates the periodicity of the oscillations along $V_{G}^\mathrm{dc}$. (\textbf{B}) Same at \SI{100}{\mega\hertz}. The vertical dashed line indicates the point where the oscillations disappear at $V_\mathrm{AWG} =$~\SI{220}{\milli\volt}.}
    \label{figS3}
\end{figure}

In order to calibrate the amplitude of the square drive $V_{G}^\mathrm{ac}$, we study the transmission of the FPI in the dc regime, i.e. where instead of sending a pulse we only apply a dc voltage on the input of the interferometer. In the meantime, we change the amplitude of the square drive $V_{G}^\mathrm{ac}$ for periodic signals at \SI{1}{\giga\hertz} (figure \ref{figS3}A) and \SI{100}{\mega\hertz} (figure \ref{figS3}B). We observe that, for the right value of the ac amplitude of the square signal $V_{G}^\mathrm{ac}$, the oscillations of the transmission vanish and transmission becomes flat as a function of $V_{G}^\mathrm{dc}$. This specific amplitude of the square drive corresponds to a $\pi$ phase shift, such that the interference pattern is averaged out between the phases $0$ and $\pi$. The amplitude of the square voltage for which this happens changes based on the frequency of the square drive, at \SI{1}{\giga\hertz} the $\pi$ shift happens at $V_G^\mathrm{ac}=$~\SI{280}{\milli\volt}, and at \SI{100}{\mega\hertz}, this shift happens at $V_G^\mathrm{ac}=$~\SI{220}{\milli\volt}. This discrepancy originates from an attenuation at \SI{1}{\giga\hertz} that is not present at \SI{100}{\mega\hertz}. At higher amplitude of the square drive, the contrast reappears with a $\pi$ phase shift of the oscillations. In figure \ref{figS3}B, we observe some jumps that can be attributed to random charge effects within the resonator.

\subsection{Analysis procedure}
\label{analysis_procedure}
\begin{figure}
    \centering
    \includegraphics[width = 0.35\textwidth]{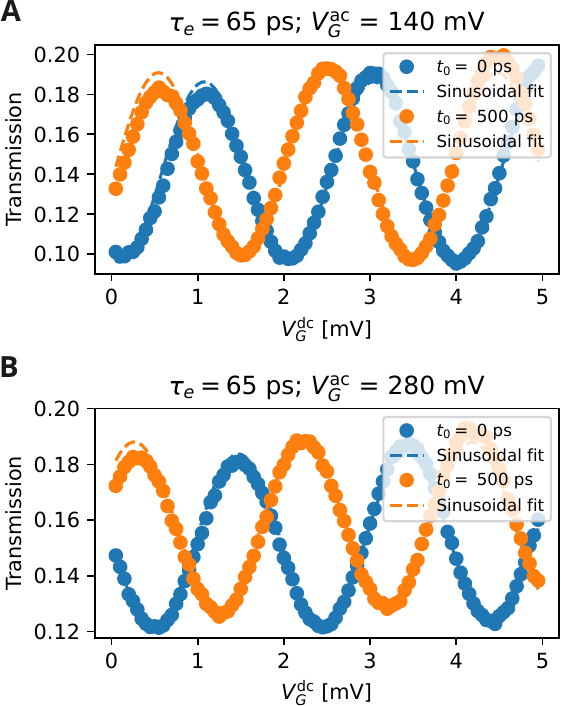}
    \caption{\textbf{Analysis procedure:} (\textbf{A}) Dependence of the transmission with $V_G^\mathrm{dc}$ for $\tau_e =$~\SI{65}{\pico\second} and $V_{G}^\mathrm{ac}=$~\SI{140}{\milli\volt} corresponding to data from figure \ref{figS2}D. The two data sets correspond to times $t_0=$~\SI{0}{\pico\second} (blue) and $t_0 = $~\SI{500}{\pico\second} (orange). They are fitted using a sinusoidal function (dashed lines). (\textbf{B}) Same for $\tau_e=$~\SI{65}{\pico\second} and $V_{G}^\mathrm{ac}=$~\SI{280}{\milli\volt}, corresponding to data from figure \ref{figS2}E.}
    \label{figS4}
\end{figure}

In order to extract the phase and amplitude of the interferometric signal, we perform cuts on the two-dimensional maps $T(V_G^\mathrm{dc},t_0)$  (shown in figure \ref{figS2}) at fixed $t_0$. These cuts show an oscillating signal as a function of $V_G^\mathrm{dc}$ which is fitted using a cos function of the form $A\cos{(V_G^\mathrm{dc}/V_0+\vartheta)}+b$ as shown in figure \ref{figS4}. The fit parameters $\vartheta$ and $A$ are then used to plot the main text figures 3G-J. The contrast is then calculated as the ratio $A(t_0)/\mathrm{max}\big(A(t_0)\big)$.

From these fits we observe that there is no second harmonic contribution to the signal and that a simple sinusoidal oscillation describes our experimental data perfectly, justifying the use of a model where a single round-trip inside the FP cavity is taken into account. 

\subsection{Contrast vs. amplitude of gate oscillations}

\begin{figure}[h!]
    \centering
    \includegraphics[width = 0.5\textwidth]{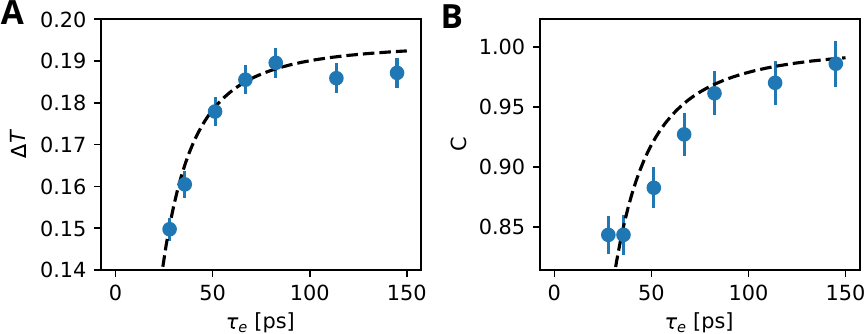}
    \caption{\textbf{Comparison between amplitude and contrast:} (\textbf{A}) Extraction of the amplitude of oscillations of $T(V_G^\mathrm{dc})$ as a function of the pulse width $\tau_e$. We extract a cavity time $\tau_L=$~\SI{30}{ps}. (\textbf{B}) Extraction of the contrast of the oscillations (taking into account the varying base line of the data) in $V_G^{\mathrm{dc}}$. The cavity time $\tau_L=$\SI{30}{\pico\second} used for the black dashed line is the same in both figures.}
    \label{figS5}
\end{figure}

In figures 2B and 2C of the main text, we present the evolution of the contrast as a function of the electronic temperature and the width of the electronic pulse. We made the choice to plot the contrast instead of the amplitude of the oscillations. However, as we show in figure \ref{figS5}, both quantities lead to the same result when plotted as a function of the pulse width. In figure \ref{figS5}A the blue dots correspond to the amplitude of the oscillations extracted, as explained in appendix \ref{analysis_procedure}. In figure \ref{figS5}B, we present the same data points as in the main text when,instead of the contrast, we plot the amplitude of the oscillations which corresponds to the amplitude divided by the base line offset of the gate dependent oscillations. The contrast is normalized to the value for $\tau_e\rightarrow\infty$. In both figures, the dashed line represents the overlap of the wave packets
\begin{equation}
    \mathrm{Re}\left[\int{\mathrm{d}t\varphi_{\tau_{e}}(t)\varphi^{*}_{\tau_e}(t+\tau_L)}\right] = \frac{1}{1+(\tau_L/2\tau_e)^2}
\end{equation}
from which we can extract the value of $\tau_L$. In figures \ref{figS5}.A and B, the dashed line only differ by a numerical factor in order to rescale this overlap to the quantity that is plotted.

\subsection{Characterization of the QPCs}

\begin{figure}[h!]
    \centering
    \includegraphics[width = 0.45\textwidth]{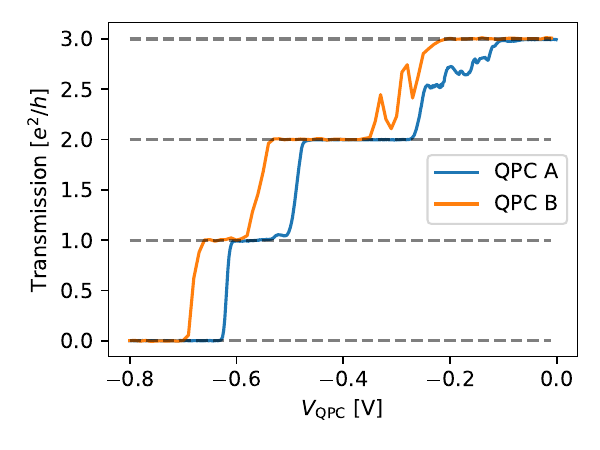}
    \caption{\textbf{Characterization of the QPCs:} Transmission measured through QPC1 and QPC2 as a function of the voltage applied on the QPCs. The dashed line represent the expected position of the quantization plateaus at the filling factor $\nu=3$.}
    \label{figS7}
\end{figure}

The data presented in the main text was acquired at the filling factor $\nu=3$. We characterize the quantization of the Hall plateaus through the transmission as a function of the voltage applied on the QPCs as presented in figure \ref{figS7}. In this figure, we observe three well defined plateaus through both QPCs, quantized at integer values of $e^2/h$ with a maximal conductance of $3e^{2}/h$, as expected for the filling factor $\nu=3$.

\section{Numerical modelization}
\subsection{Modelization of the square drive}

\begin{figure}[h!]
    \centering
    \includegraphics[width = 0.4\textwidth]{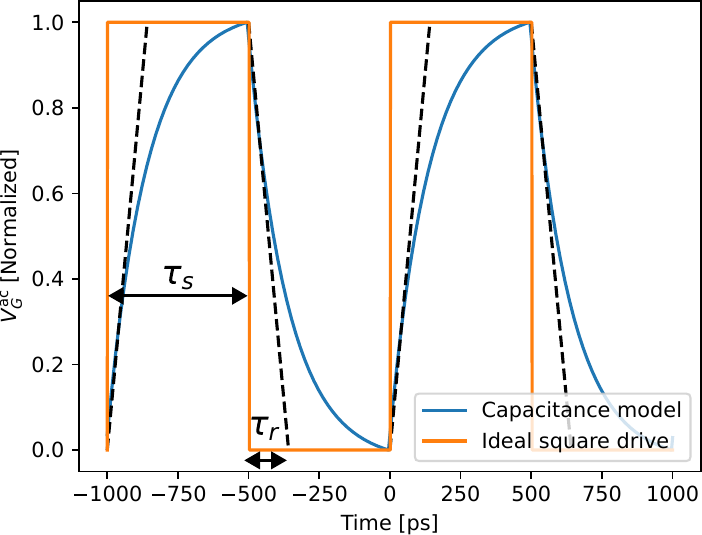}
    \caption{\textbf{Pulse modelization:} An ideal square drive (in orange) can be defined by its periodicity $\tau_S$. Due to various attenuation sources on the rf line used to apply this drive, we expect a deformation of the pulse characterized by a rise time $\tau_r=$~\SI{140}{\pico\second}.}
    \label{figS9}
\end{figure}

The periodicity of the drive applied to the gate is $2\tau_s =$~\SI{1}{\nano\second}. The ideal shape of the drive as sent by the AWG is a perfect square signal, such as the one represented on figure \ref{figS9} in orange. However, attenuation and capacitances of the cables will deform the signal that is probed at the level of the FPI. We model this signal in our simulation as an exponentially increasing and decreasing wave such as represented in figure \ref{figS9}. This exponential rise is characterized by a rise time $\tau_r$ that we take to be equal to \SI{140}{\pico\second} to better reproduce our experimental data. 

\subsection{Continuous evaluation of the parameters}

\begin{figure}[h!]
    \centering
    \includegraphics[width = 0.45\textwidth]{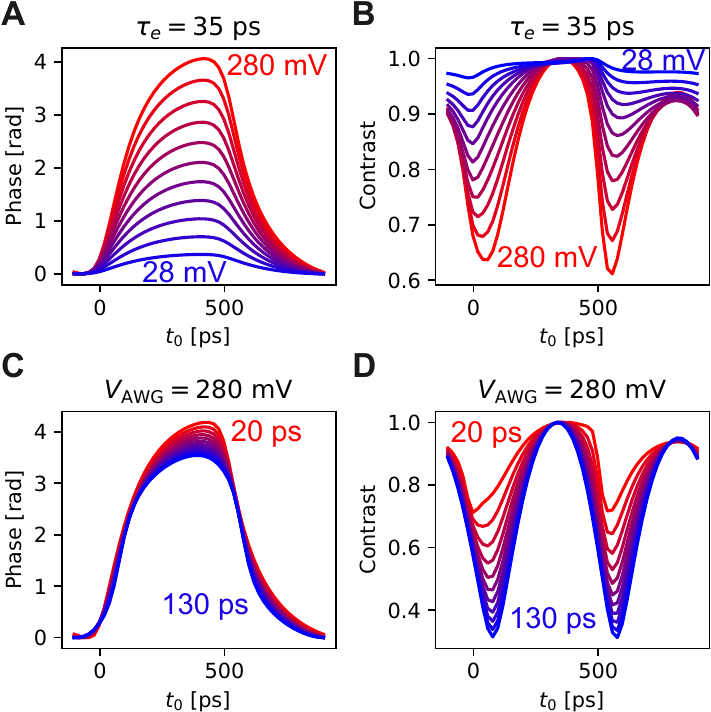}
    \caption{\textbf{Continuous evolution of the parameters:} (\textbf{A}) Evolution of the phase calculated  for $\tau_e=$~\SI{35}{\pico\second} for voltages $V_\mathrm{AWG}$ continuously evolving from \SI{28}{\milli\volt} (in blue) to \SI{280}{\milli\volt} (in red). (\textbf{B}) Associated contrast. (\textbf{C}) Evolution of the phase calculated for $V_\mathrm{AWG}=$~\SI{280}{\milli\volt} for pulse widths evolving continuously from \SI{20}{\pico\second} (red) to \SI{130}{\pico\second} (blue). (\textbf{D}) Associated contrast.}
    \label{figS10}
\end{figure}

Figure \ref{figS10} presents the result of the evaluation of the transmission for parameters $V_\mathrm{AWG}$ and $\tau_e$ varying continuously over a larger range than the one shown in the main text. Figures \ref{figS10}.A. and B. respectively show the phase and contrast for $\tau_e=$~\SI{35}{\pico\second} and amplitudes of the square drive evolving from \SI{28}{\milli\volt} (blue) to \SI{280}{\milli\volt} (red). We can see on panel A that the phase varies linearly with the amplitude of the square drive. Regarding the contrast, the dips around $t_0=0$ and $t_0=$~\SI{500}{\pico\second} become more and more pronounced when the amplitude of the drive increases. As discussed in the  main text, the dips of the contrast are related to quantum fluctuations of the phase that increase for increasing amplitude of the square. 

Figures \ref{figS10}C and D respectively show the phase and contrast for $V_\mathrm{AWG}=$\SI{280}{\milli\volt} for continuously varying pulse widths from \SI{20}{\pico\second} (red) to \SI{130}{\pico\second} (blue). As we have discussed in the main text, increasing the width of the pulse leads to a reduced phase shift compared to the expected variation for a given amplitude of the square. As seen on panel D, increasing the temporal width of the single electron wavepackets also increases the quantum fluctuations of the phase, leading to more pronounced dips of the contrast at $t_0 \approx 0$ and $t_0 \approx$~\SI{500}{\pico\second}.  

\bibliography{biblio}

\end{document}